\documentclass[a4paper,11pt]{article}
\pdfoutput=1 

\usepackage{jcappub}
\usepackage{graphicx}
\usepackage{dcolumn}
\usepackage{amssymb,amsmath,bm}
\usepackage{color}
\usepackage[dvipsnames,table]{xcolor}
\usepackage{xfrac}
\usepackage{aas_macros}
\usepackage{mathrsfs}
\usepackage{subcaption}
\usepackage{rotating}
\usepackage{chngcntr}
\usepackage{units}
\usepackage{multirow}
\usepackage{xspace}
\usepackage{xfrac}
\usepackage{mathtools}
\usepackage{comment}
\usepackage{soul}
\usepackage{gensymb}
\usepackage{multirow}

\newcommand{\cN}{\mathcal{N}}
\newcommand{\Om}{\Omega_{\rm m}}

\newcommand{\Ob}{\Omega_{\rm b}}
\newcommand{\ns}{n_{\rm s}}
\newcommand{\mpch}{$h^{-1}$Mpc}

\newcommand{\lenangle}[1]{\left\langle #1 \right\rangle}
\newcommand{\mean}[1]{\bar{#1}}

\newcommand{\lmax}{\ell_{\rm max}}

\newcommand{\fsky}{f_{\rm sky}}
\newcommand{\Apix}{A_{\rm pix}}

\newcommand{\nside}{N_{\rm side}}
\newcommand{\nv}{\hat{\bf n}}

\newcommand{\shear}{\boldsymbol{\gamma}}
\newcommand{\ellip}{\boldsymbol{e}}

\newcommand{\cM}{\mathcal{M}} 
\newcommand{\cF}{\mathcal{F}} 

\newcommand{\lcdm}{$\Lambda$CDM\xspace}
\newcommand{\cmbk}{CMB$\kappa$\xspace}
\newcommand{\nmt}{\texttt{NaMaster}\xspace}
\newcommand{\healpix}{\texttt{HEALPix}\xspace}
\newcommand{\ccl}{\texttt{CCL}\xspace}
\newcommand{\class}{\texttt{CLASS}\xspace}

\newcommand{\hfit}{\textsc{HALOFIT}\xspace}
\newcommand{\mpyth}{\texttt{MontePython}\xspace}

\newcommand{\des}{DES\xspace}
\newcommand{\desgc}{DES$g$\xspace}
\newcommand{\deswl}{DES$\shear$\xspace}
\newcommand{\kids}{KiDS\xspace}
\newcommand{\planck}{{\sl Planck}\xspace}
\newcommand{\eboss}{eBOSS-QSO\xspace}
\newcommand{\redmagic}{\textsc{redMaGiC}\xspace}
\newcommand{\mcal}{\textsc{Metacalibration}\xspace}
\newcommand{\dls}{DELS\xspace}
\newcommand{\northd}{ND\xspace}
\newcommand{\southd}{SD\xspace}
\newcommand{\alld}{FD\xspace}

\usepackage{natbib}
\title{The growth of density perturbations in the last $\sim$10 billion years from tomographic large-scale structure data}

\author[a,b,c]{Carlos Garc\'ia-Garc\'ia,}
\author[a]{Jaime Ruiz Zapatero,}
\author[a]{David Alonso,}
\author[e,a]{Emilio Bellini,}
\author[a]{Pedro G. Ferreira,}
\author[a]{Eva-Maria Mueller,}
\author[d]{Andrina Nicola,}
\author[b,c]{Pilar Ruiz-Lapuente}

\affiliation[a]{Department of Physics, University of Oxford, Denys Wilkinson Building, Keble Road, Oxford OX1 3RH, United Kingdom}
\affiliation[b]{Instituto de F\'isica Fundamental, Consejo Superior de Investigaciones Cient\'ificas, c/. Serrano 123, E-28006, Madrid, Spain}
\affiliation[c]{Institut de Ci\`{e}ncies del Cosmos (ICCUB), c/. Mart\'i i Franqu\'es 1, E-08028, Barcelona, Spain}
\affiliation[d]{Department of Astrophysical Sciences, Princeton University, Peyton Hall, Princeton, NJ 08544, USA}
\affiliation[e]{Département de Physique Théorique, Université de Genève, 24 quai Ernest Ansermet, 1211 Genève 4, Switzerland}

\emailAdd{carlos.garcia-garcia@physics.ox.ac.uk}

\abstract{In order to investigate the origin of the ongoing tension between the amplitude of matter fluctuations measured by weak lensing experiments at low redshifts and the value inferred from the cosmic microwave background anisotropies, we reconstruct the evolution of this amplitude from $z\sim2$ using existing large-scale structure data. To do so, we decouple the linear growth of density inhomogeneities from the background expansion, and constrain its redshift dependence making use of a combination of 6 different data sets, including cosmic shear, galaxy clustering and CMB lensing. We analyze these data under a consistent harmonic-space angular power spectrum-based pipeline. We show that current data constrain the amplitude of fluctuations mostly in the range $0.2<z<0.7$, where it is lower than predicted by \planck. This difference is mostly driven by current cosmic shear data, although the growth histories reconstructed from different data combinations are consistent with each other, and we find no evidence of systematic deviations in any particular experiment. In spite of the tension with \planck, the data are well-described by the \lcdm model, albeit with a lower value of $S_8\equiv\sigma_8(\Om/0.3)^{0.5}$. As part of our analysis, we find constraints on this parameter of $S_8=0.7781\pm0.0094$ (68\% confidence level), reaching almost percent-level errors comparable with CMB measurements, and 3.4$\sigma$ away from the value found by \planck.}

\begin{document}

\maketitle

  \section{Introduction}\label{sec:intro}
    The current era of precision cosmology has seen a tremendous growth in the volume, quality, and variety of astronomical data that have become available. This has allowed us to progressively improve the constraints on the cosmological parameters  of the $\Lambda$ Cold Dark Matter (\lcdm) model. The \lcdm model is a remarkable fit to most observations, ranging from the cosmic microwave background (CMB) \cite{1807.06209} to large scale structure (LSS) inferred from galaxy clustering and weak lensing \cite{1708.01530,2007.08991,2007.15632}, yet  a number of intriguing discrepancies or ``tensions'' have begun to arise.
    A particularly important tension is related to the amplitude of density fluctuations (or linear matter perturbations) at low redshifts predicted by CMB data in comparison with direct measurements by cosmic shear and galaxy clustering data. This is commonly summarized in the so-called $S_8$ parameter, defined as
    \begin{equation}
      S_8\equiv\sigma_8\,\left(\frac{\Om}{0.3}\right)^\alpha,
    \end{equation}
    where $\sigma_8$ is the variance of the linear matter overdensity field in spheres with a 8 \mpch{} radius, and $\Om$ is the fractional energy density of non-relativistic matter, both defined at $z=0$. The exponent $\alpha$ is chosen to minimize the correlation between $S_8$ and $\Om$, although the choice $\alpha=0.5$, which we use here, is a good approximation in most cases. This quantity has been measured by various weak lensing surveys \cite{1708.01538,1809.09148,2007.15632} , encountering different levels of tension with the value inferred from measurements by the \planck satellite. Overall, cosmic shear data tend to recover somewhat smaller values of $S_8$ than the CMB and, to date, the strongest disagreement has been reported by the Kilo-Degree Survey collaboration (\kids) \cite{2007.15632}, with a significance around $3\sigma$.

    State-of-the-art measurements now exploit the complementarity between weak gravitational lensing and galaxy clustering~\cite{1708.01530,2007.15632,2105.03421}. Weak lensing is a direct tracer of the integrated density perturbations between source and observer, and therefore can be used to measure their amplitude in a relatively clean way. Its cumulative nature, however, washes out most features in the distribution of these perturbations, and makes it difficult to track the evolution of the amplitude in detail~\cite{astro-ph/9912508,1706.09359}. Galaxy clustering, on the other hand, is a high signal-to-noise, but biased, tracer of the local (as opposed to integrated) density perturbations. The galaxy bias makes it difficult to extract information about the growth of structure from the projected galaxy distribution alone, although important features in this distribution (e.g. baryon acoustic oscillations) can still be recovered~\cite{1705.05442, 1712.06209}. In combination with weak lensing data, however, galaxy clustering improves our ability to reconstruct the history of density perturbations, significantly enhancing the associated cosmological constraints \cite{astro-ph/9912508,1706.09359,1708.01530,1710.03235,2007.15632,2105.03421}.

    This paper focuses on reconstructing the growth history, i.e. the evolution of the amplitude of density perturbations (characterized by $\sigma_8$ or $S_8$) as a function of redshift. The motivation for this endeavour is twofold: on the one hand, reconstructing this evolution from existing data sets will allow us to understand the redshift ranges over which current data are able to constrain the growth of structure, potentially shedding light on the origin of the aforementioned ``$S_8$ tension''. On the other hand, the amplitude of density perturbations is, arguably, the natural ``observable'' to which projected weak lensing surveys are sensitive, in the same way that the density-weighted growth rate, $f\sigma_8$, is for redshift-space distortions and the transverse and longitudinal distance indicators, $\alpha_\perp$ and $\alpha_\parallel$, are for baryonic acoustic oscillations (BAOs)  in spectroscopic clustering data sets\footnote{The case of photometric LSS data is less clear-cut than that of spectroscopic surveys, since weak lensing and projected galaxy clustering are in principle sensitive to both growth and geometry \cite{2010.05924,2105.09545}. Nevertheless, combined shear-clustering analyses constitute a unique probe of structure growth.} \cite{1979Natur.281..358A,1987MNRAS.227....1K,2007.08991}. Thus, this analysis will allow us to present current constraints from photometric redshift surveys in terms of observables, as opposed to final parameter constraints, which can then be used to e.g. study deviations with respect to \lcdm.

    To do this, we will make use of various galaxy clustering and weak lensing data sets. In particular, we will use galaxy clustering and cosmic shear data from the first data release of the Dark Energy Survey (\des, \cite{1708.01530}), cosmic shear data from the fourth data release of the \kids collaboration \citep{2007.01845}, galaxy clustering from the DESI Legacy Survey \citep{1804.08657,2010.00466}, the clustering of high-redshift quasars in the extended Baryon Oscillation Spectroscopic Survey \citep{2007.08999}, and maps of the CMB weak lensing convergence made by the \planck collaboration \cite{1807.06210}. 
    These data will allow us to recover the growth history in the range $0.2 \lesssim z \lesssim 2$, as well as to compare the histories reconstructed by the two different cosmic shear experiments (\des and \kids) independently, and by the combination of clustering and CMB lensing in the absence of shear data.

    A key aspect of this work is that our analysis will be based on an independent estimation of the relevant two-point correlations between data sets using a consistent harmonic-space-based framework to estimate both angular power spectra and their covariance matrix. This has the added value of being able to compare the constraints obtained by the different experiments, as well as their combination, under the same analysis pipeline (including the modelling of systematic effects). As we shall show, in some cases this leads to a significant improvement (in terms of goodness of fit, for example) with respect to previous analyses.

    For the impatient reader, our main findings are:
    \begin{itemize}
      \item Decoupling the linear growth of fluctuations from the background expansion, and modelling the former through quadratic splines, we show that existing large-scale structure data prefer a lower amplitude of fluctuations in the range $0.2\lesssim z\lesssim0.7$ than that predicted by \planck by $\sim5\%$. This is also the range of redshifts where current data are most sensitive to structure growth.
      \item This result is recovered consistently by independent data set combinations, and is driven by existing cosmic shear data.
      \item The recovered growth history is in good agreement with a \lcdm model, although with a lower value of $S_8$ than that predicted from CMB. From a combined analysis of all the data, we obtain a constraint $S_8=0.7781\pm0.0094$. This is in $\sim3.4\sigma$ tension with current CMB data and with smaller uncertainties.
    \end{itemize}

    This work is organized as follows. Section \ref{sec:theory} describes the theoretical background used to model the different auto- and cross-correlations, as well as the method used to parametrize the growth history. Section \ref{sec:data} describes the different data sets used in our analysis, together with the methods used to process them into maps tracing the projected mater overdensities. In Section \ref{sec:meth} we present the methods used to estimate all power spectra and their covariance matrices, and the likelihood used to connect data and theory. Section \ref{sec:res} presents the main results of this analysis in terms of parameter constraints on \lcdm and on the growth history. The main results are then discussed and summarized in Section \ref{sec:disc}.

  \section{Theory}\label{sec:theory}
    \subsection{Projected anisotropies}\label{ssec:theory.signal}
      Our analysis will be based on a set of cross-correlations between fields defined on the celestial sphere $u(\nv)$ that are related to a three-dimensional quantity $U({\bf x},z)$ in the lightcone through line-of-sight integrals with a radial kernel $q_u(\chi)$ \cite{astro-ph/9912508}:
      \begin{equation}\label{eq:projection}
        u(\nv)=\int d\chi\,q_u(\chi)\,U(\chi\nv,z(\chi)),
      \end{equation}
      where $\chi$ is the comoving radial distance. In the Limber approximation \cite{1953ApJ...117..134L}, appropriate for the broad kernels used here, the angular power spectrum between two projected fields $u$ and $v$ is related to the three-dimensional power spectrum of their associated quantities $U$ and $V$ via:
      \begin{equation}\label{eq:limber}
        C^{uv}_\ell=\int\frac{d\chi}{\chi^2}q_u(\chi)q_v(\chi)\,P_{UV}\left(k=\frac{\ell+1/2}{\chi},z(\chi)\right).
      \end{equation}

      Our analysis will consider three fields: the projected galaxy overdensity $\delta_g$, the galaxy shear $\gamma_G$, and the CMB lensing convergence $\kappa$:
      \begin{itemize}
        \item $\delta_g$ is related to the three-dimensional galaxy overdensity $\Delta_g$ via a radial kernel proportional to the redshift distribution of sources in the tomographic bin. Assuming a simple linear bias model relating $\Delta_g$ and the 3D matter overdensity $\Delta_M$, the effective radial kernel for galaxy clustering is
        \begin{equation}
          q_{\delta_g}(\chi)=b_gp(z)\frac{dz}{d\chi},
        \end{equation}
        where $b_g$ is the linear galaxy bias, and $p(z)$ is the redshift distribution normalized to unit integral.

        In the case of eBOSS quasars, we will also take into account the impact of lensing magnification on the observed clustering. Magnification is caused by a combination of the displacement in galaxy positions and the modification in the observed source flux due to gravitational lensing. It is thus a direct tracer of the matter overdensity $\Delta_M$ with a radial kernel
        \begin{equation}
           q_{\mu_g}=-(2-5s)K_\ell\,q_L(\chi),
        \end{equation}
        where $s$ is the slope of the source magnitude distribution, and $q_L$ is the lensing kernel
        \begin{equation}
          q_L(\chi)\equiv\frac{3}{2}H_0^2\Om\frac{\chi}{a(\chi)}\int_{z(\chi)}^\infty dz' p(z')\frac{\chi(z')-\chi}{\chi(z')}.
        \end{equation}
        $a\equiv1/(1+z)$ is the scale factor, and $K_\ell$ is a scale-dependent factor that accounts for the difference between the three-dimensional transverse Laplacian, connecting the matter overdensity and the gravitational potential, and the angular Laplacian on the sphere:
        \begin{equation}
          K_\ell\equiv\frac{\ell(\ell+1)}{(\ell+1/2)^2},
        \end{equation}
        which is only significantly (above $\sim1\%$) different from 1 on large scales ($\ell\lesssim10$). We will use $s=0.2$ for the quasar sample \cite{astro-ph/0504510}. The final kernel for the eBOSS quasars is given by the sum of $q_{\delta_g}$ and $q_{\mu_g}$.
        \item $\gamma_G$ and $\kappa$ are directly related to the three-dimensional matter overdensity $\Delta_M$ with radial kernels
        \begin{equation}
           q_{\gamma_G}(\chi)=G_\ell\,q_L(\chi),\hspace{12pt}
           q_{\kappa}=K_\ell\frac{3}{2}H_0^2\Om\frac{\chi}{a(\chi)}\frac{\chi_*-\chi}{\chi_*},
        \end{equation}
        where
        \begin{equation}
          G_\ell\equiv\sqrt{\frac{(\ell+2)!}{(\ell-2)!}}\frac{1}{(\ell+1/2)^2}
        \end{equation}
        is the equivalent of $K_\ell$ for the shear field, and $\chi_*$ is the comoving distance of the source plane (i.e. the distance to the last scattering surface in the case of CMB lensing). We will also account for the impact of intrinsic alignments on the galaxy shear field, using the non-linear alignment model of \cite{astro-ph/0406275,0705.0166}. In this model, the intrinsic alignment contribution to the observed shear, $\gamma_I$, is proportional to the local tidal field, and therefore its radial kernel is proportional to the redshift distribution of the source sample:
        \begin{equation}
          q_{\gamma_I}(\chi)=-G_\ell A_{\rm IA}(z)p(z)\frac{dz}{d\chi}.
        \end{equation}
        We will parametrize the intrinsic alignment amplitude $A(z)$ as was done for the analysis of the DES first-year data \cite{1708.01530, 1708.01538}:
        \begin{equation}\label{eq:ias}
          A_{\rm IA}(z)=A_{{\rm IA},0}\left( \frac{1+z}{1+z_0} \right)^{\eta_\text{IA}}\frac{0.0139 \Om}{D(z)}\,,
        \end{equation}
        with $A_{\text{IA},0}$ and $\eta_{\text{IA}}$ two free parameters, $z_0$ the redshift pivot (which we fix to $z_0=0.62$ as in \cite{1708.01530, 1708.01538}), and $D(z)$ the linear growth factor. As in the case of the eBOSS quasars, the final kernel is given by the sum of $q_{\gamma_G}$ (or $q_{\kappa}$) and $q_{\gamma_I}$.  In our analysis we will also account for residual multiplicative biases in the shear power spectra. These enter the power spectra as an overall multiplicative factor $(1+m)$ for each shear field being correlated, with $m$ a free parameter of the model.
      \end{itemize}

      As described above, the three tracers (galaxy clustering, cosmic shear and CMB lensing) used here can be used to measure the matter density fluctuations. The last remaining ingredient of the model therefore is the matter power spectrum $P_{mm}(k, z)$ entering Eq. \ref{eq:limber}. The model used here is described in the next section. We used the Core Cosmology Library\footnote{\url{https://github.com/LSSTDESC/CCL}} (\ccl) \cite{1812.05995} to compute all cosmological quantities, making use of the \class~\cite{1104.2933} Boltzmann code to calculate the linear matter power spectrum.

    \subsection{Growth reconstruction}\label{ssec:theory.growth_rec}
      Given the ongoing debate around the value of the amplitude of matter fluctuations at late time measured by large-scale structure data~\cite{1708.01530, 1708.01538, 2007.15632, 1809.09148, 2105.09545} compared with that extrapolated by CMB data in the context of \lcdm ~\cite{1807.06209,2011.11613}, our main objective in this paper is to explore: a) whether this tension can be directly mapped into the time evolution of inhomogeneities within the range of redshifts covered by existing data sets, and b) whether the growth histories recovered by different data sets are compatible with each other. To do so, we will decouple the linear growth factor $D(z)$ from the background cosmological parameters and instead treat it as a free function that we will constrain directly from the data. 
      
      In a $\Lambda$CDM Universe with no massive neutrinos, in which the only relevant density inhomogeneities are those of pressureless matter, the linear matter overdensity field $\Delta^{\rm L}_M$ grows in a ``self-similar'' fashion, in which time dependence is factorizable:
      \begin{equation}\label{eq:gf_factor}
        \Delta_M^{\rm L}({\bf k},z)=D(z)\Delta_M^L({\bf k},0).
      \end{equation}
      $D(z)$ is the linear growth factor normalized as $D(0)=1$. This result is easily understandable since, in this case, the equation for the evolution of matter overdensities (conservation of the energy momentum tensor) is scale independent. This factorizability then maps directly into the linear matter power spectrum:
      \begin{equation}\label{eq:dz_selfsim}
        P_{\rm L}(k,z)=D^2(z)\,P_L(k,0).
      \end{equation}

      The dependence of the growth factor on redshift as a time variable can be a rich observable to constrain the main energy components of the Universe. Unfortunately, the non-linear nature of the gravitational interaction causes the matter overdensity to quickly depart from this linear behaviour, causing smaller, non-linear scales to grow faster than larger, linear ones. Fortunately, for a wide range of cosmological models, the power spectrum of the observable non-linear overdensity can be expressed as a functional of the linear power spectrum:
      \begin{equation}
        P_{\rm NL}(k,z)=F[P_{\rm L}](k,z),
      \end{equation}
      with little dependence on the specific ingredients of the cosmological model beyond those that determine the form of $P_{\rm L}$ and its evolution in time \citep{astro-ph/9705121,astro-ph/0202358,astro-ph/0505565,astro-ph/0610213,1208.2701,1606.05345}. Here we will make use of one such parametrization to connect the linear and non-linear power spectra: the popular \hfit model as implemented in \cite{1208.2701}.

      There are different options to parametrize $D(z)$, such as expanding it as a linear combination of basis functions (e.g. as a polynomial in $z$ or $a$) or, more ideally, modelling it as a Gaussian process with its hyperparameters and conditional distribution determined from the data. Existing off-the-shelf parameter inference frameworks for the cosmological analysis of large-scale structure data are not efficient enough yet to deal with the high-dimensional parameter spaces associated with Gaussian processes (although the community is moving fast in that direction \cite{2104.14568,2105.02256}), and therefore we choose a middle ground. In our case, $D(z)$ is determined by its value at a set of fixed redshift nodes $\tilde{D}_{z_i}\equiv D(z_i)$. Each node is treated as an additional free parameter in the likelihood. The growth factor $D(z)$ is then calculated at any $z$ by interpolating over the values of $\tilde{D}_{z_i}$, extrapolating beyond the range covered by the fixed redshift nodes $z_i$. Specifically, we use a quadratic spline interpolation in the space $(\log(1+z),\log(D(z)))$. In order to avoid allowing for unphysically large or negative values of $D(z)$ at high redshifts where our data have no constraining power, we fix $D(z)$ to the $\Lambda$CDM prediction with the best-fit \planck cosmological parameters beyond $z=5$\footnote{We verified that this choice does not affect the constraints recovered on $D(z)$ in the range where we have data by comparing our results with those found by fixing $D(a = 10^{-4})$ to its fiducial value instead.}.

      We choose four redshift nodes centered at the mean redshifts of some of the galaxy clustering tracers used in our analysis. The logic behind this choice is that, on large scales, a combination of galaxy clustering and weak lensing data can be used to effectively measure the galaxy bias and the amplitude of matter fluctuations at the mean redshift of the clustering sample, and thus the galaxy clustering bins act as natural anchors at which growth is effectively measured. In particular, we choose the position of the first, third and fifth redshift bins for the \des clustering sample (see Section \ref{sssec:data.DES.gc}), as well as the mean redshift of the \eboss sample. The nodes are thus located at $z_i\in\{0.24, 0.53, 0.83, 1.5\}$. As stated above, this is in addition to a fixed node at $z=5$ that matches \planck's best fit \lcdm cosmology.

      There are a few caveats associated with the method chosen to quantify structure growth. First, the extrapolation above and below the redshift range covered by the nodes may lead to biases in the final constraints. This is more relevant in the range $z<0.24$ which the cosmic shear data are sensitive to, rather than at high redshifts. Additionally, the choice of a quadratic spline instead of other interpolation methods will likely have an impact on the final constraints on the spline parameters (e.g. in terms of their final uncertainty or correlation between modes). As we discuss in Section~\ref{ssec:res.growth}, neither of these effects has a strong impact on our final result. Additionally, the self-similar linear growth assumed here (Eq. \ref{eq:gf_factor}) is not valid in general (e.g. in the presence of massive neutrinos, or through scale-dependent growth in modified gravity models e.g.~\cite{Deffayet:2010qz}). Current constraints on neutrino mass from particle physics experiments and CMB observations place any signature of neutrinos below the level of detectability of the data and scales used here, and therefore any significant departure from \lcdm in terms of growth history we observe is unlikely to be due to the scale-dependent signature of massive neutrinos. Finally, the use of \hfit to relate the linear and non-linear power spectra, while sufficiently accurate for a wide family of cosmological models, is not guaranteed to be appropriate for models with an arbitrary growth history. Since the deviations with respect to \lcdm found in our analysis are relatively mild, we do not expect this to have a significant impact on our conclusions.

  \section{Data}\label{sec:data}
    \begin{figure}
      \centering
      \includegraphics[width=0.9\textwidth]{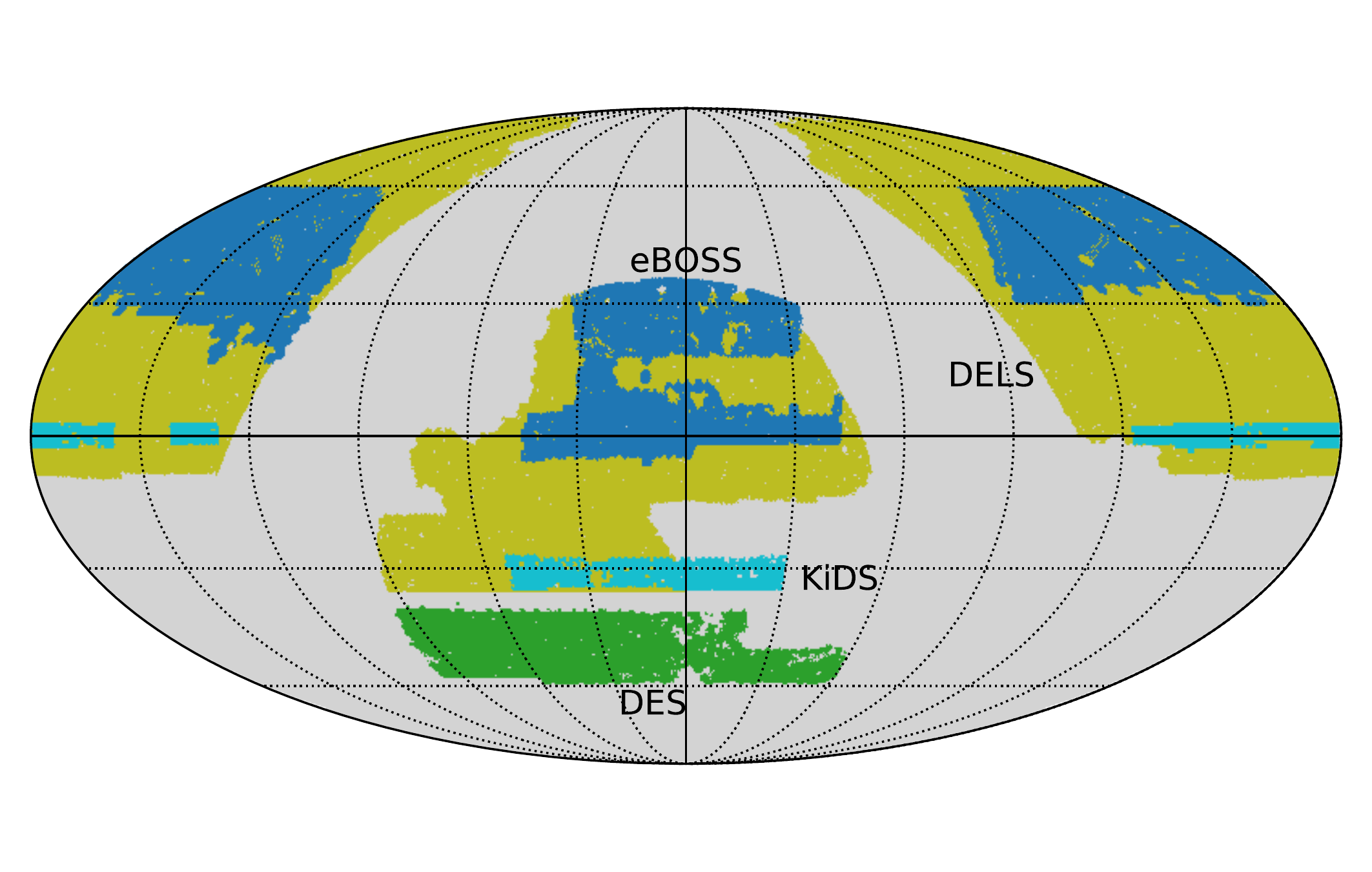}
      \caption{Sky footprint of the galaxy surveys used in this analysis. The \planck \cmbk map overlaps with all the data sets and is not shown. We carry out separate analyses of the \des (green) and \kids$+$\dls (cyan and yellow) $3\times2$pt data, which we label \southd and \northd respectively. When analysed separately, we include their combination with \planck \cmbk and \eboss (dark blue). We will also consider the combination of all the data sets shown.}
      \label{fig:footprint}
    \end{figure}
    This work is based on the analysis of 6 different data sets. These are: the cosmic shear and galaxy clustering samples used in the cosmological analysis of the first-year data release of the Dark Energy Survey (DES$g$ and DES$\gamma$ respectively, \cite{1708.01530}, Section \ref{ssec:data.DES}), the cosmic shear sample used in the fourth data release of the Kilo-Degree Survey (KiDS1000, \cite{2007.15633}, Section \ref{ssec:data.KiDS}), a galaxy clustering sample extracted from the DESI Legacy Survey for the analysis presented in \cite{2010.00466} (\dls, Section \ref{ssec:data.dls}),  the clustering of quasars in the extended Baryon Oscillation Spectroscopic Survey (\eboss, \cite{2007.08999}, Section \ref{ssec:data.eBOSS}), and the lensing convergence of the CMB measured by \planck (\cmbk, \citep{1807.06210}, Section \ref{ssec:data.Planck}).

    Our work will conceptually divide these data sets into two groups of data, determined by the data combinations for which a ``$3\times2$-point'' analysis (i.e. the combination of two-point correlations involving cosmic shear and galaxy clustering) can be carried out over non-overlapping sky regions. First, the ``South data set'' (\southd) will comprise the DESY1 galaxy clustering and cosmic shear samples. The ``North data set'' (\northd) instead focuses on the combination of cosmic shear from KiDS1000 and galaxy clustering from \dls. The \northd and \southd sets also include the cross-correlation of their clustering and shear samples with the \planck convergence map in their respective footprints. Finally, when analyzed separately, both data sets also include auto-correlations of the eBOSS-QSO sample, and its cross-correlations with the CMB lensing map, in order to provide a high-redshift lever arm for the growth reconstruction. We will also consider the combination of all six data sets (which we will label \alld). The specific auto- and cross-correlations between the different probes and their associated scale cuts are described in Section \ref{ssec:meth.pcl}.

\begin{figure}
      \centering
      \includegraphics[width=0.99\textwidth]{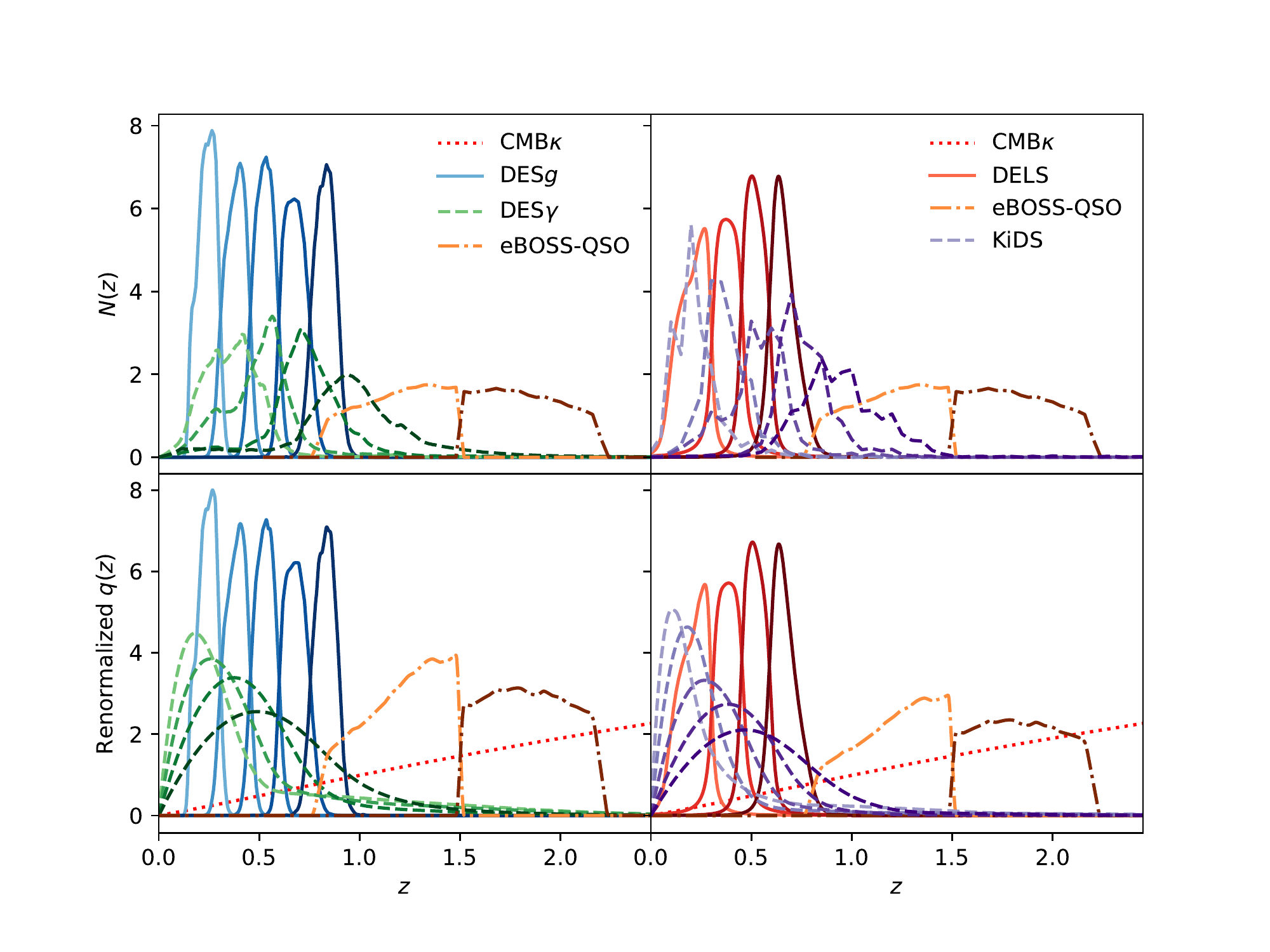}
      \caption{Redshift distributions (top) and radial kernels (bottom) for the different data sets used in this work: \southd (left) and \northd (right). All curves are scaled by an arbitrary normalizing factor to make them visible on the same scale. The cosmic shear kernels are weighted towards smaller redshifts than those covered by the sources, while the clustering kernels are local in redshift.}\label{fig:zbins}
    \end{figure}

    Figure~\ref{fig:footprint} shows the sky footprints covered by each of the data sets used here. The redshift distributions and associated radial kernels for each tracer used in our analysis (as defined in Eq. \ref{eq:projection}) are shown in Figure \ref{fig:zbins}. In combination, the data used in this analysis allow us to cover the range of redshifts $z\lesssim2$ as well as a significant fraction of the celestial sphere.

    The next subsections describe the procedure used to process these data sets and extract four key data products: signal maps, sky masks, noise power spectra and redshift distributions. A summary is provided in Table \ref{tab:data_products}.
    \begin{table}[]
      \small
      \centering
      \begin{tabular}{|l|c|c|c|}
      \hline
      {\bf Tracer} & {\bf Signal map} & {\bf Mask} & {\bf Noise power spectrum} \\
      \hline
      DES$g$ & Eq. \ref{eq:des.gc.signal} & PDR &  Eq. \ref{eq:des.gc.nl} \\
      \hline
      DES$\gamma$ & Eq. \ref{eq:des.sh.signal} (unit weights) & Eq. \ref{eq:des.sh.mask} (unit weights) & Eq. \ref{eq:des.sh.noise} (unit weights)\\
      \hline
      eBOSS-QSO & Eq. \ref{eq:eboss.signal} & Random counts map & High-$\ell$ average \\
      \hline
      Planck$\kappa$ & PDR & PDR $+$ 0.2$^\circ$ apodization & PDR \\
      \hline
      \dls & Eq. \ref{eq:des.gc.signal} (unit weights) & PDR \cite{2010.00466} & Eq. \ref{eq:des.gc.nl} (unit weights)\\
      \hline
      \kids & Eq. \ref{eq:des.sh.signal} & Eq. \ref{eq:des.sh.mask} & Eq. \ref{eq:des.sh.noise}\\
      \hline
      \end{tabular}
      \caption{Summary table describing the methods used to generate the main map-level data products needed to estimate power spectra and covariance matrices (signal maps, masks and noise power spectra) for the 6 data sets used in this analysis. The cells reading ``PDR'' correspond to quantities that are directly provided in the public data release associated with these data.}\label{tab:data_products}
    \end{table}

    \subsection{DES Y1 galaxy clustering and weak lensing}\label{ssec:data.DES}
      The Dark Energy Survey is a 5 year survey that will cover $\unit[5000]{deg^2}$ in 5 filter bands (\textit{grizY}) and has mapped hundreds of millions of galaxies and thousands of galaxy clusters \cite{astro-ph/0510346}. These observations are taken from the Cerro Tololo Inter-American Observatory (CTIO) with the $\unit[4]{m}$ Blanco Telescope, using the 570-megapixel Dark Energy Camera (DECam~\cite{1504.02900}). In this paper we use the first year of data products\footnote{The DESY1 data are available at \url{https://des.ncsa.illinois.edu/releases/y1a1}.}, which cover $\unit[1786]{deg^2}$ before masking \cite{1801.03181,1708.01531}.  In this analysis we use the same fiducial galaxy samples used in the DES Y1 $3\times2$pt analysis \cite{1708.01530}, including their associated redshift distributions. Our fiducial analysis will also employ the same models used in \cite{1708.01530} to describe various systematic effects, as discussed in Section \ref{ssec:theory.signal}.

      \subsubsection{Galaxy clustering}\label{sssec:data.DES.gc}
        We use the clustering sample presented in \cite{1708.01536}. The sample was constructed using the \redmagic algorithm, which selects red luminous galaxies with excellent photometric redshift accuracy ($\sigma_z = 0.017(1+z)$ in this case, \cite{1507.05460}). We split this sample into the same five redshift bins used in the analysis of \cite{1708.01536,1708.01536, 1708.01530}. The three lower redshift bins are populated with galaxies from the \redmagic high-density sample. These have a comoving density $\mean{\rho}_{\rm{n}} \simeq 10^{-3}$ and minimum luminosity, $L_{\text{min}} \simeq 0.5 L_*$. The other two redshift bins contain galaxies from the \redmagic high-luminosity ($\mean{\rho}_{\rm{n}} \simeq 4\times 10^{-4}$, $L_{\text{min}} = L_*$) and higher-luminosity ($\mean{\rho}_{\rm{n}} = 10^{-4}$, $L_{\text{min}} = 1.5 L_*$) samples respectively. 

        To track the survey geometry, we use the mask publicly available with the Y1 release, in the form of a \healpix~\footnote{\url{http://healpix.sourceforge.net/}} \cite{astro-ph/0409513} map at resolution $N_{\rm side}=4096$ containing the effective fractional area $w_p$ of each pixel $p$. In order to avoid inaccuracies from strongly masked pixels, we set the value of all pixels with $w_p<0.5$ to zero. This leaves a total unmasked area of $\unit[1321]{deg^2}$.

        All galaxies in the \redmagic tomographic bins are assigned weights to correct for the impact of observational systematics. In order to create the associated overdensity map for each redshift bin we must therefore use both these weights and the geometric information contained in the sky mask. To do so we first create a map containing the weighted number of galaxies lying in a given pixel $p$, $n_p = \sum_{i\in p} v_i$, where $v_i$ the weight of the $i$-th galaxy. The overdensity field is then given by
        \begin{equation}\label{eq:des.gc.signal}
          \delta_{p} = \frac{n_p}{\mean{n} w_p} - 1,
        \end{equation}
        where  $\mean{n} = \sum_p n_p / \sum_p w_p$ is the weighted average galaxy number per pixel and $w_p$ the value of the mask on that pixel.

        The auto-correlation galaxy clustering power spectra has a shot noise contribution, $N_\ell$, that has to be subtracted. Assuming perfect Poisson sampling, this can be done analytically \cite{1912.08209}. In short, the ``mode-coupled'' noise power spectrum (i.e. before multiplying by the inverse mode-coupling matrix in Eq. \ref{eq:pcl_mcm}) is given by
        \begin{equation}\label{eq:des.gc.nl}
          \tilde{N}_\ell = \frac{\langle w\rangle}{\mean{n}_2},
        \end{equation}
        where $\langle w\rangle$ is the mean value of the mask across the full sky, and $\mean{n}_2$ is the effective mean number density, given by
        \begin{equation}
          \mean{n}_2\equiv\frac{(\sum_i v_i)^2}{A_{\rm pix}\sum_p w_p\sum_i v_i^2}.
        \end{equation}
        Here, $A_{\rm pix}$ is the pixel area in steradians. Note that $\mean{n}_2$ reduces to $\mean{n}$ in the case of equal weights.

        The DES Y1 release provides an estimate of the redshift distribution for the five clustering redshift bins. We use these in our analysis marginalizing over a parameter corresponding to the mean of each distribution.

      \subsubsection{Galaxy shear}\label{sssec:data.DES.sh}
        The DES shear analysis was carried out using two different shape-measurement algorithms, \textsc{IM3SHAPE}~\cite{1302.0183} and \mcal~\cite{1702.02601}. In this work we make use of the \mcal catalog. \mcal fits a 2D Gaussian model for each galaxy to the pixel data in the $r$, $i$ and $z$ bands, convolved with their corresponding point-spread function (PSF). This process is repeated with artificially sheared images to calibrate the shear estimator, which allows for the calculation of shear-dependent selection effects that could bias the statistics a few percents \cite{1708.01533,1708.01537,1708.01538,1702.02601}.
        
        The raw ellipticities measured by \mcal, $\hat{\ellip}_i$, must be corrected as
        \begin{equation}\label{eq:des:ecal}
          \ellip_i=\hat{\ellip}_i/\bar{R},
        \end{equation}
        where $\ellip_i$ is the $i$-th component of the resulting calibrated ellipticity. The multiplicative correction is $\bar{R}=({\sf R}_{11}+{\sf R}_{22})/2$, where ${\sf R}$ is the $2\times 2$ response tensor calculated by \mcal. The response tensor contains two additive terms: the shear response ${\sf R}_{\shear}$, obtained from artificially sheared images as described above, and the selection response ${\sf R}_{\rm S}$, which accounts for the selection bias that appears when applying a set of selection criteria on the sheared galaxy sample. The total response tensor is ${\sf R}={\sf R}_{\shear}+{\sf R}_{\rm S}$ which is almost diagonal with ${\sf R}_{11} \sim {\sf R}_{22}$, and thus its effect can be well approximated by $\bar{R}$. Note that \mcal computes the shear response tensor ${\sf R}_{\shear}$ for each galaxy but, as done in \cite{1708.01538,1708.01533}, we average it over the whole sample in each redshift bin to calculate $\bar{R}$. Finally, the DES analysis found a non-negligible mean residual ellipticity ($\bar{\ellip}_i\sim\mathcal{O}(10^{-4})$) in each redshift bin. Following \cite{1708.01538}, we subtracted this mean ellipticity per redshift bin after calibration.

        We generate maps of the two shear components $\shear_{1,2}$ as the per-pixel weighted average of galaxy ellipticities
        \begin{equation}\label{eq:des.sh.signal}
          \shear_{i,p} = \frac{\sum_{n\in p} v_n \ellip_{n,i}}{\sum_n v_n} ,
        \end{equation}
        where $i=1,2$ is the shear component, $v_n$ is the weight associated with the $n$-th galaxy. \mcal assigns unit weights to all galaxies $v_n = 1$ in it. As discussed in \cite{2010.09717}, we use the sum of weights in each pixel as the mask associated with the resulting shear maps. This should be a close-to-optimal choice assuming the weights are close to inverse-variance:
        \begin{equation}\label{eq:des.sh.mask}
          w_p = \sum_{i \in p} v_i\,,
        \end{equation}

        The mode-coupled noise power spectrum is calculated analytically as \cite{2010.09717}:
        \begin{equation}
          \widetilde{N}_{\ell > 2} = \Apix \lenangle{\sum_{i\in p} v_i^2 \sigma^2_{e,i}}_{\rm pix},\label{eq:des.sh.noise}
        \end{equation}
        where $\sigma^2_{e,i} = (e_{i,1}^2 + e_{i,2}^2)/2$ is an estimate of the shape-measurement noise rms per galaxy. As shown in \cite{2010.09717}, this estimate is equivalent to averaging over the power spectra of a large number of catalogs with randomly rotated ellipticities, at a much lower computational cost. This approach is exact in the noise-dominated regime, where the contribution from the cosmic shear signal to each galaxy is negligible. Note that, as a spin-2 field, the power spectrum of the shear field is zero at $\ell<2$.

        The \mcal sample is divided into 4 tomographic redshift bins. We use the fiducial redshift distributions for each bin provided with the Y1 data release. These were estimated by stacking the per-galaxy probability distributions derived by the BPZ \cite{astro-ph/0605262} photo-$z$ algorithm for all sources in a given bin \cite{1708.01532}. They were further validated by cross-matching against the COSMOS 30-band catalog \cite{1604.02350}, and via cross-correlations (see \cite{1708.01532} for further details). To a large extent, most of the uncertainty on the redshift distribution for cosmic shear samples can be well-described by an uncertainty in the mean of said distribution, since a shift in this mean impacts the width of the associated redshift kernel significantly \cite{2003.11558}. Thus, as done in \cite{1708.01538}, we marginalize over four parameters corresponding to linear shifts in the mean of the redshift distributions of each bin.

    \subsection{eBOSS quasars}\label{ssec:data.eBOSS}
      In order to extend the range of redshifts over which we reconstruct the growth history, we use quasar (QSO) clustering measurements from the extended Baryon Oscillation Spectroscopic Survey (eBOSS) derived the from Sloan Digital Sky Survey Data Release 16 (DR16). In particular we use the homogeneous quasar sample used for the cosmological power spectrum analysis of \cite{2007.08999,2007.08998}, and presented in \citep{2007.09000}. The catalog comprises 343,708 objects with measured redshifts in the range $0.8 \leq z \leq 2.2$, covering over 4,800 deg$^2$. Note that we will not use the existing redshift-space distortion measurements from \eboss to constrain growth here. Instead, we use the quasar sample as another two-dimensional projected tracer of the large-scale structure which we correlate with all the other datasets.

      The footprint of this sample is described by a set of random sources, covering the same area in the absence of clustering. Furthermore, in order to correct for the modulation in the observed number density of objects caused by various systematics, observing conditions and Galactic systematics, objects in both the random and data catalogs are assigned weights. Since we carry out a projected 2D analysis, we include all systematic weights (accounting for redshift failures, fiber collisions, sky systematics), but omit the so-called ``FKP'' weights that maximize the signal-to-noise ratio of the three-dimensional power spectrum.

      We combine eBOSS measurements from the North and South Galactic Caps into one single catalog that we split into two different bins with redshifts above and below $z=1.5$ respectively. The redshift distribution of each bin is estimated directly from the data as a histogram of the measured spectroscopic redshifts. We do not account for any systematic uncertainty in the redshift distribution thus constructed.

      In order to calculate the quasar overdensity map we make use of the random catalog to track the survey geometry. Specifically, the overdensity in pixel $p$ is calculated as
      \begin{equation}\label{eq:eboss.signal}
        \delta_p = \frac{\sum_{i \in p} v_{d,i} - \alpha \sum_{j \in p}v_{r,j}}{w_p},
      \end{equation}
      where $v_{d,i}$ and $v_{r,i}$ are the data and random weights for the $i$-th object lying in pixel $p$. $\alpha = \sum_p v_{d, p} / \sum_p v_{r, p}$ accounts for the fact that the random catalog is significantly larger than the data to minimize the impact of its associated shot noise. $w_p$ is the survey mask, which we compute as the scaled sum of random weights in each pixel:
      \begin{equation}
        w_p = \alpha \sum_{i \in p} v_{r, i}.
      \end{equation}
      In order to minimize the impact of shot noise in the random catalog when constructing this mask, we do so at a relatively low resolution ($N_{\rm side}=512$), ensuring a sufficiently high average number density of random points in non-empty pixels. This is then upgraded to our target resolution ($N_{\rm side}=4096$) correcting for the different pixel area. 

      The \eboss catalog is by far the sparsest of the clustering samples used in this analysis, and the angular quasar power spectrum is dominated by shot noise over a large range of scales. A careful treatment of the noise bias in the auto-correlation is therefore crucial in order to obtain reliable constraints from it. This is further complicated by the fact that we need to account for the impact of shot noise in the random catalog, which affects both the numerator and denominator of Eq. \ref{eq:eboss.signal}. Fortunately, since the auto-correlations are noise-dominated on small scales, we can obtain a reasonable estimate of the noise bias from the data. An uncorrelated noise component would appear in the mode-coupled power spectrum as a scale-independent contribution, which we can thus estimate by averaging the value of the power spectrum calculated from the masked overdensity map in the range $2000 \leq \ell < 2 \nside = 8192$. We find this method to be a better approximation than analytically calculating the noise contribution via e.g. a generalization of Eq. \ref{eq:des.gc.nl} accounting for data and random weights. Nevertheless, given the importance of the shot-noise contribution, we additionally marginalize over a constant noise power spectrum with a free amplitude and 10\% Gaussian prior centered on the power spectrum amplitude estimated as we describe in Section \ref{ssec:meth.like}.

    \subsection{CMB lensing from \planck}\label{ssec:data.Planck}
      The CMB weak lensing convergence field (\cmbk) is produced by the matter overdensities between the last scattering surface and us, with most of the contribution coming from  $0.5 \lesssim z \lesssim 3$ \citep{astro-ph/9810257,astro-ph/0601594}. In this work, we use the convergence map made available as part of the \planck 2018 data release \cite{1807.06210}. The map covers a sky fraction $\fsky = 0.671$ and thus overlaps spatially with all surveys considered here. Various photometric redshift surveys have made use of cross-correlations with CMB lensing maps from different collaborations to extract cosmological constraints \cite{1810.02322,2008.04369,2010.00466,2011.11613,2102.07701,2103.15862,2105.03421}.

      Planck is a third-generation space mission, following COBE and WMAP, dedicated to measure the CMB anistropies, offering full sky maps of temperature and polarization anisotropies with micro-Kelvin sensitivity per resolution element. The specifications of the 2018 \planck CMB lensing data release are described in detail in \cite{1807.06210}.

      In this analysis we make use of the ``Minimum Variance'' (MV) lensing convergence harmonic coefficients, which we transform into a \healpix map at $N_{\rm side}=4096$ resolution. The harmonic coefficients are provided in the range $\ell<4096$. Following \cite{1502.01591, 1807.06210} we remove the smallest multipoles $\ell < 8$, which are too sensitive to the mean-field subtraction in the lensing reconstruction process,  and go up to $\ell = 2000$, falling inside the \textit{aggressive} scale ranges of \cite{1807.06210}. We use the binary sky mask made available with this map. Since the lensing reconstruction noise power spectrum rises sharply with $\ell$ on small scales, we apodize this mask with a $0.2^\circ$ ``C1'' kernel \cite{1809.09603} in order to minimize the leakage from noise-dominated small-scale modes. The final usable sky fraction is $\fsky \simeq 0.66$.
      
      Our analysis does not include the CMB lensing auto-correlation, and therefore we do not need to include a rigorous modelling of the various noise bias terms that enter the CMB lensing likelihood. The CMB lensing noise, however, enters the covariance matrix for any power spectra involving the CMB lensing map. For this, we use the estimate of the noise power spectrum $N_\ell^{\kappa}$ provided with the data release.

    \subsection{KiDS-1000 weak lensing}\label{ssec:data.KiDS}
      The Kilo Degree Survey (\kids) is a large optical survey that has mapped $\unit[1350]{deg^2}$ of the sky in four tomographic bands ($ugri$) using the VLT Survey Telescope (VST) located in the ESO Paranal Observatory. Its main objective is the measurement of the cosmic shear signal. In this paper, we use  the Gold Sample from the public data release 4 (DR4) \cite{1902.11265}. This data release includes both images covering a total area of $\unit[1006]{deg^2}$, which reduces to $\unit[777.4]{deg^2}$ after masking, as well as forced photometry data in five additional infrared bands from the VIKING survey. The Gold Sample is described in \cite{1902.11265,2007.01845} and it targeted a sample of galaxies with reliable shapes and redshift distributions.
      
      We follow the analysis of \cite{2007.15633}. We split the gold sample in 5 tomographic bins assigning each galaxy to a bin based on its best-fitting photometric redshift, $z_B$. As described in \cite{2007.01845}, in each redshift bin we subtract the residual weighted mean ellipticity, and correct for the multiplicative bias factors listed in Table 1 of \cite{2007.15633}, estimated by \kids from image simulations. We then produce shear maps and weight maps (mask) for each redshift bin following Eqs. \ref{eq:des.sh.signal} and \ref{eq:des.sh.mask} respectively, making use of the shear measurement weights assigned by the {\sl lens}fit algorithm. The coupled noise power spectrum is estimated analytically as in Eq. \ref{eq:des.sh.noise}.

      We use the redshift distributions provided with the DR4. These redshift distributions were constructed using the self-organizing map (SOM) method described in \cite{1909.09632,2007.15635}. As in the case of \des, we will marginalize over the mean of each redshift distribution in the likelihood.

    \subsection{The DESI Legacy Imaging Survey}\label{ssec:data.dls}
      The DESI Legacy Imaging Survey (\dls) \cite{1804.08657}, meant to locate targets for the Dark Energy Spectroscopic Instrument (DESI, \cite{1308.0847}), combines photometry from three different telescopes: the DECam Legacy Survey \cite{1504.02900,2016AAS...22831701B} at declinations $\delta<33^\circ$, the Mayall $z$-band Legacy Survey \cite{10.1117/12.2231488,2016AAS...22831702S}, and the Beijing-Arizona Sky Survey \cite{10.1117/12.552189,1908.07099}. The final survey covers 17,739 deg$^2$.
      
      Our analysis will make use of the galaxy sample selected by \cite{2010.00466} to obtain cosmological constraints from its cross-correlation with CMB temperature and lensing convergence\footnote{The data can be found at \url{https://gitlab.com/qianjunhang/desi-legacy-survey-cross-correlations}. We thank the authors for the remarkable care with which these data were made publicly available.}. We also follow their analysis choices closely. Each galaxy was assigned a redshift based on a multi-dimensional matching in colour space with a set of spectroscopic samples. We use these redshifts to separate the sample into the four tomographic bins used in \cite{2010.00466}.

      Our main objective in using these data is to combine with the \kids cosmic shear data to carry out a 3$\times$2pt analysis that can then be compared with the results found with the \des Y1 data. In order to facilitate this comparison, as well as the combination of both data sets, we remove all data with declination $\delta<-36^\circ$ from the \dls sample, ensuring no area overlap between \dls$+$\kids and \des.

      For each redshift bin, we compute a first estimate of the galaxy overdensity map using Eq. \ref{eq:des.gc.signal}, where $n_p$ is the number of galaxies in pixel $p$, $\bar{n}$ is the mean number density, and $w_p$ is the completeness of the pixel (understood as its effective fractional area). $\bar{n}$ is estimated as $\bar{n}=\sum_{p\in G}n_p/\sum_{p\in G}w_p$, where $G$ is the set of ``good'' pixels with completeness above 95\% and a star density lower than $N_{\rm star} = 8515$ deg$^{-2}$ \citep{2010.00466}. We make use of the completeness map and star map made available by \cite{2010.00466}. The latter corresponds to a smoothed version of the ALLWISE total density map \citep{1805.11525}. Additionally, all pixels with completeness $w_p < 0.86$ or star density $N_{\rm star} > \unit[1.29 \times 10^4]{deg^{-2}}$ were masked. The overdensity field thus created contains residual star contamination, affecting the galaxy auto-correlation on large scales. We correct for this at the map level by subtracting a systematic overdensity map whose pixel values are estimated by evaluating a 5th-order polynomial fit to the mean galaxy overdensity with respect to the local logarithmic number of stars in each pixel, as done in \cite{2010.00466}. As in the case of \des, we estimate the coupled noise power spectrum through Eq. \ref{eq:des.gc.nl}.

      We model the redshift distribution of each tomographic bin using the same approach described in \cite{2010.00466}. The true redshift distribution can be related to the photo-$z$ distribution through a convolution with the conditional distribution $p(z_t|z_p)$, where $z_t$ and $z_p$ are the true and photometric redshifts respectively. This conditional distribution is modeled to be stationary (i.e. only dependent on $z_t-z_p$) and given by a Lorentzian distribution of the form
      \begin{equation} \label{eq:Lorentzian}
        L(\delta z) = \frac{N}{(1 + ((\delta z - x_0) / \sigma^2)^2 / 2a)^a}.
      \end{equation}
      The parameters $x_0$, $\sigma$ and $a$ are determined from the existing spectroscopic data and given in Table 1 of \cite{2010.00466}. Additionally, the authors of \cite{2010.00466} marginalized over $x_0$ and $a$ in their cosmological analysis, showing that they can be self-calibrated to a large extent through the use of clustering cross-correlations between different redshift bins. In order to keep a consistent model among different probes, we simply marginalize over linear shifts in the mean of each redshift distribution, and fix $x_0$ and $a$ to the values found by \cite{2010.00466} after self-calibration.

  \section{Methods}\label{sec:meth}
    \subsection{Power spectra}\label{ssec:meth.pcl}
      We analyze all data using a common harmonic-space power spectrum framework based on the pseudo-$C_\ell$ method as implemented in \nmt \cite{1809.09603}, including the approximations described in \cite{1906.11765,2010.09717} to estimate the power spectrum covariance. We describe the method briefly below and direct the reader to these references for further details on the implementation\footnote{The analysis pipeline used to process the public data and estimate all power spectra and covariances can be found in \url{https://github.com/xC-ell/growth-history/}, together with the data products and plots of the power spectra.}.

      An observed map $\tilde{a}$ is modelled as the product of the true map $a$ and a weights map $w$:
      \begin{equation}
        \tilde{a}(\nv) = w(\nv) a(\nv).
      \end{equation}
      Although, for simplicity, we will limit our discussion to scalar (spin-0) fields such as $\delta_g$ or $\kappa$, the methods used are directly generalized to spin-$2$ (or, in fact arbitrary spin) fields such as $\shear$. This is described in detail in \cite{1809.09603,1906.11765,2010.09717}.  The weights map can be understood both as a mask, i.e., a map defining whether a given pixel has been observed or not ($w=1$ and 0 respectively), as well as an inverse-variance local weight, down-weighting regions of high noise.

     Following the convolution theorem, the spherical harmonic coefficients of $\tilde{a}$ are a convolution of the spherical harmonic coefficients of $a$ and $w$, a fact that leads to mode coupling between the power spectra of the observed and true fields. Defining the ``coupled pseudo-$C_\ell$'' $\tilde{C}^{ab}_\ell$ between fields $a$ and $b$ with weight maps $v$ and $w$ respectively as
      \begin{equation}
        \tilde{C}^{ab}_{\ell}=\frac{1}{2\ell+1}\sum_{m=-\ell}^\ell \tilde{a}_{\ell m}\tilde{b}^*_{\ell m},
      \end{equation}
      its relation to the true underlying $C^{ab}_\ell$ is given by
      \begin{equation}\label{eq:pcl_mcm}
        \langle\tilde{C}^{ab}_\ell\rangle=\sum_{\ell'}M^{vw}_{\ell\ell'}C^{ab}_{\ell'},
      \end{equation}
      where $\langle\,\,\rangle$ denotes averaging over realizations of $a$ and $b$. $M^{vw}_{\ell\ell'}$ is the so-called \emph{mode-coupling matrix} (MCM), and depends solely on the weight maps of the two fields being correlated. As shown in \cite{astro-ph/0105302}, $M^{vw}_{\ell\ell'}$ can be computed efficiently and analytically thanks to the orthogonality of the Wigner $3j$ symbols. Explicit expressions for the coupling matrix of different combinations of spin-0 and 2 fields can be found in \cite{1809.09603}. In the limit of full-sky data, $M^{vw}_{\ell\ell'}$ is the identity, and departures from this limit give rise to a statistical off-diagonal coupling between neighbouring multipoles. In many practical cases the MCM is non-invertible, and thus $\tilde{C}^{ab}_\ell$ cannot be turned into an unbiased estimator of $C_\ell^{ab}$. The pseudo-$C_\ell$ estimator then proceeds along the following three steps:
      \begin{enumerate}
        \item \ul{Binning}. As a way to regularize the MCM, we bin the power spectra into \emph{bandpowers} containing weighted sums of different $\ell$s:
        \begin{equation}
          \tilde{C}^{ab}_q=\sum_{\ell\in q}B_q^\ell \tilde{C}^{ab}_\ell.
        \end{equation}
        We can then define the \emph{binned MCM}
        \begin{equation}
          \cM^{vw}_{qq'}\equiv\sum_{\ell\in q}\sum_{\ell'\in q'}B_q^\ell M^{vw}_{\ell\ell'},
        \end{equation}
        which is usually invertible for sufficiently broad bandpowers.
        \item \ul{Inverting the MCM.} The decoupled bandpowers $\hat{C}^{ab}_q$ are then defined by inverting the binned MCM:
        \begin{equation}
          \hat{C}_q^{ab} = \sum_{q'} (\cM^{vw})^{-1}_{q q'} (\tilde{C}_{q'}^{ab} - \tilde{N}^{ab}_{q'})\,,
        \end{equation}
        where we have explicitly removed the (binned) mode-coupled noise power spectrum $\tilde{N}^{ab}_q$. The form of $\tilde{N}^{ab}_q$ depends on the maps being correlated, and their calculation has been described in Section \ref{sec:data} for the different data sets used here.
        \item \ul{Bandpower convolution.} $\hat{C}^{ab}_q$ would be an unbiased estimator of the true power spectrum evaluated at, e.g., the central multipole of each bandpower, if the latter was exactly constant within each bandpower. Since this is not the case, the effects of binning must be propagated through, in order to connect the observed bandpowers with a given theoretical prediction\footnote{It is important to note that this is not a source of bias in the pseudo-$C_\ell$ estimator. The resulting bandpower window functions are independent of the underlying power spectrum and depend only on the structure of the mode-coupling matrix.}. The theoretical prediction for the bandpowers $C^{ab}_q$ is related to the theory power spectrum $C^{ab}_\ell$ by convolving the latter with the \emph{bandpower window functions} $\cF^{vw}_{q\ell}$ through a fast matrix-vector multiplication
        \begin{equation}
          C_q^{ab} = \sum_\ell \cF^{vw}_{q \ell} C^{ab}_\ell,
        \end{equation}
        where
        \begin{equation}
          \cF^{vw}_{q \ell} = \sum_{q'} (\cM^{vw})^{-1}_{q q'} \sum_{\ell' \in q'} w^{\ell'}_q M^{uv}_{\ell' \ell}\,.
        \end{equation}
        $\cF^{ab}$ encodes the three pseudo-$C_\ell$ linear operations: mode-coupling, binning, and binned mode-decoupling.
      \end{enumerate}

      We use a common \healpix resolution of $N_{\rm side}=4{,}096$ for all sky maps, which allows us to compute all power spectra up to a maximum multipole $\ell_{\rm max}=3N_{\rm side}-1=12{,}287$. We bin all power spectra into a common set of bandpowers with the following binning scheme: we use linear bins with width $\Delta \ell = 30$ between $0 \leq \ell \leq 240$, and logarithmic bins until $\lmax$ with $\Delta \log_{10}(\ell) = 0.055$. The resulting bandpower edges are listed in Table 2 of \cite{2010.09717}.

    \subsection{Covariance matrix}\label{ssec:meth.cov}
      The power spectrum covariance matrix can be decomposed in three contributions \cite{0810.4170,1302.6994}: the disconnected ``Gaussian'' part, equivalent to assuming that all fields involved are Gaussianly distributed, the connected non-Gaussian part (cNG), corresponding to the intrinsic connected trispectrum of the fields, and the super-sample covariance (SSC), caused by the coupling of small-scale modes sourced by modes larger than the survey footprint. As we show in appendix \ref{app:cov}, the cNG and SSC contributions are subdominant on the scales used here, and our fiducial analysis will employ only the Gaussian contribution.
      
      The Gaussian covariance can be computed analytically and depends on the survey geometry, which causes statistical correlations between different bandpowers. Accounting for this effect is particularly critical for the case of cosmic shear, given the high complexity of its associated weights map (which follows the distribution of the source galaxies). However, the exact calculation of these correlations sales as $O(\lmax^6)$, making it intractable for our data. In this work, instead, we use the Narrow Kernel Approximation (NKA) presented in~\cite{astro-ph/0307515,1906.11765}, and improved in \cite{2010.09717} for the case of cosmic shear. The method scales as $O(\lmax^3)$ and has been shown to be accurate for all spin-0 and spin-2 quantities used here. The core assumption behind the NKA is that the map-level mode-coupling matrix is close to diagonal, such that the power spectra can be treated as constant in the calculation. This is usually an excellent approximation for well-behaved masks, but fails catastrophically in the case of cosmic shear, or for steep power spectra. As shown in \cite{2010.09717}, the NKA can be improved significantly in these cases if one simply substitutes the power spectra used in the calculation by their mode-coupled versions scaled by the overlapping sky fraction:
      \begin{equation}\label{eq:clcov}
          C^{ab}_\ell\hspace{6pt}\rightarrow\hspace{6pt}\frac{\tilde{C}^{ab}_\ell}{\langle v\,w\rangle_{\rm pix}}=\frac{1}{\langle v\,w\rangle_{\rm pix}}\sum_{\ell '}M^{vw}_{\ell\ell'}C^{ab}_{\ell'}.
      \end{equation}

      If the noise properties vary across the footprint, the effective masks of the signal and noise components of the maps are not the same, and thus the signal-signal, signal-noise and noise-noise contributions to the covariance matrix should in principle be computed using different mode-coupling coefficients \cite{1811.05714}. As shown in \cite{2010.09717}, this additional complication can be avoided by simply adding to the signal power spectrum (Eq. \ref{eq:clcov}), the equivalent coupled noise power spectrum scaled by the overlapping sky fraction. We have described the calculation of the coupled noise power spectrum for the different data sets used here in Section \ref{sec:data}. While this is not an exact result, the impact of this approximation on the covariance matrix is negligible for our current analysis (although its validity must be reassessed in for future, more sensitive cosmic shear data sets). As a technical note, we find that the ``spin-0'' approximation discussed in \cite{1906.11765}, which treats the $E$ and $B$ components as independent scalar fields, yields a better estimate of the shear covariance matrix for the \kids data, given its significant depth variations across the footprint, and thus we make use of this approximation here. In addition to this, we make use of the ``Toeplitz approximation'' proposed in \cite{2010.14344} to accelerate the calculation of the mode-coupling coefficients used in the covariance calculation. The impact of this approximation is at the sub-percent level for our data.

      This calculation of the covariance matrix requires an estimate of the power spectra of the different fields. We calculate these using \ccl assuming the \planck 2018 best-fit cosmological parameters:
      \begin{equation}\label{eq:fiducial}
        (\Omega_c,\Omega_b,h,n_s,\sigma_8)=(0.2640,0.0493,0.6736,0.9649,0.8111),
      \end{equation}
      as well as the following values for the linear galaxy bias of the \des \redmagic sample $b_g^{\rm RM}=(1.48,1.76,1.78,2.19,2.23)$, the \dls sample $b_g^{\rm DELS}=(1.13,1.40,1.35,1.77)$, and the \eboss sample $b_g^{\rm QSO}=(2.1,2.5)$. We verified that these parameters are a reasonable fit to the measured power spectra on the scales used in this analysis. For the \cmbk auto-correlation, we use the signal$+$noise power spectra provided with the public \planck data release.
    
    \subsection{Likelihood}\label{ssec:meth.like}
      \begin{table}
        \centering
        \def\arraystretch{1.2}
        \begin{tabular}{|cc|}
        \hline
        \multicolumn{2}{|c|}{\textbf{Cosmological parameters priors}} \\
        \hline
        Parameter &  Prior\\  
        \hline 
        $\Om$  &  $U (0.1, 0.9)$                             \\ 
        $A_\mathrm{s}/10^{-9}$ &  $U (0.5, 5.0)$             \\
        $\ns$ & $U (0.87, 1.07)$                             \\
        $\Ob$  &  $U (0.03, 0.07)$                           \\
        $h_0$   &  $U (0.55, 0.91)$                          \\
        $\tau$  & 0.08                                       \\
        $\sum m_\nu$ & 0                                     \\ 
        \hline
        \multicolumn{2}{|c|}{\textbf{Growth reconstruction}} \\
        $\sigma_8^{\rm fid}$ &    $0.8111$             \\
        $\tilde{D_{z_i}}$ & $U(0, 2.5)$        \\
        $\tilde{D_5}$ & 0.2116\\
        \hline
        \end{tabular}
        \caption{Prior distributions for the cosmological parameters. When reconstructing the growth history, we fix the normalization of the linear power spectrum template to $\sigma_8^{\rm fid} = 0.8111$ and do not vary $A_{\rm s}$. This is taken into account in our growth reconstruction described in Section~\ref{ssec:theory.growth_rec}. Furthermore, we fix the highest redshift node, $\tilde{D_5}$, at its value for the best fit of \planck (Eq.~\ref{eq:fiducial}), ensuring that we recover our fiducial growth at $z\geq 5$. $U(a, b)$ and $\cN(\mu, \sigma)$ are a uniform distribution with boundaries $(a, b)$, and a Gaussian distribution with mean $\mu$ and variance $\sigma$, respectively. The priors largely follow the choices of \cite{1810.02322}. }\label{tab:priors_cosmo}
      \end{table}

      We use a Gaussian power spectrum likelihood to derive constraints on the free parameters describing the measured power spectra. To do so we made use of the \mpyth sampler \cite{1210.7183,1804.07261}, which we modified to interact with \ccl as the main code for theory calculations\footnote{The likelihood and the modified \texttt{MontePython} are publicly available and can be found at~\href{https://github.com/carlosggarcia/montepython_public/tree/emilio/montepython/likelihoods/cl_cross_corr_v3}{this url}.}. We sample with the Metropolis-Hasting algorithm~\cite{Metropolis, Hastings} and address the chains convergence requiring the Gelman-Rubin parameter $R-1 \lesssim 0.01$~\cite{GelmanRubin}.
      
      We consider two types of cosmological models:
      \begin{itemize}
        \item {\bf \lcdm.} Characterized by 5 free cosmological parameters: $A_s$, $n_s$, $\Om$, $\Omega_b$ and $h$; i.e. the amplitude of scalar perturbations, the primordial spectral index, the present value of the matter and baryonic density parameters, and the dimensionless Hubble parameter, respectively. The priors used for each parameter mostly follow the choices made by the \des collaboration \cite{1810.02322}, and are listed in Table \ref{tab:priors_cosmo}. Four particular choices must be noted. We fix the optical depth $\tau=0.08$, since our data are insensitive to its value, and the redshift of the last scattering surface to a fiducial value $z_* = 1100$. We impose a flat prior on $A_s$, as opposed to other common choices such as $\log A_s$, $\sigma_8$ or $S_8$. The impact of this choice on the final constraints on $S_8$ is discussed in detail in \cite{1906.09262}. Finally, we consider only cosmologies with massless neutrinos. Since these data are not able to place strong constraints on the sum of neutrino masses, this mostly allows us to accelerate the calculation of the matter power spectrum. Although these choices may have an effect on our final constraints, at the level of a few fractions of a $\sigma$, we emphasize that our \lcdm constraints are mainly aimed at validating our analysis pipeline by comparing them with those found in the literature for subsets of our data, rather than performing a thorough study of \lcdm and its extensions, which we leave for future work. 
        \item {\bf Growth reconstruction.} In this case we retain 4 \lcdm cosmological parameters: $\Om$, $\Omega_b$, $h$ and $n_s$. The amplitude of matter fluctuations is defined by the free growth factor parameters $\tilde{D}_z$ (see Section \ref{ssec:theory.growth_rec}), which define the value of the linear growth factors at redshifts $z=0.24$, 0.53, 0.83, and 1.5. Note that the overall normalization of the $\tilde{D}_z$ parameters is degenerate with that of the fiducial linear power spectrum at $z=0$ used in Eq. \ref{eq:dz_selfsim}. Thus, when generating $P_L(k,0)$ in that equation, we fix its normalization to the best-fit value of $\sigma_8$ measured by \planck, $\sigma_8^{\rm fid}=0.8111$. The actual value of $\sigma_8$ for that particular model is then given by $\sigma_8=D(0)\sigma_8^{\rm fid}$. We are able to constrain the value of all the $\tilde{D}_z$ nodes for most of the data combinations explored. However, in order to avoid unphysical (e.g. negative) values of $D_i$ in cases where some of the parameters are unconstrained (e.g. in the absence of high-redshift data), we impose a flat prior on the growth parameters $\tilde{D}_z\in[0,2.5]$. This prior is uninformative and sufficiently broad not to bias our result when the data are able to constrain these parameters.
      \end{itemize}

      Our model also contains a large number of free nuisance parameters characterizing different sources of astrophysical uncertainties and systematics. These are (see Table~\ref{tab:priors_nuisance} for their priors):
      
      \begin{table}
        \centering
        \def\arraystretch{1.2}
        \begin{tabular}{|cc|cc|}
        \hline
        \multicolumn{4}{|c|}{\textbf{Nuiscance parameters priors}} \\
        \hline
        Parameter &  Prior & Parameter &  Prior\\  
        \hline 
        \multicolumn{2}{|c|}{\textbf{DES lens photo-z bias}}  &        \multicolumn{2}{c|}{\textbf{eBOSS QSO bias}} \\
        $\Delta z_{{\rm g}}^1 $ & $\cN (0.0, 0.007)$ &             $b_\mathrm{g}^{i}$   & $U (0.8, 5.0)$ \\ \cline{3-4}
        $\Delta z_{{\rm g}}^2 $  & $\cN (0.0, 0.007)$ &            \multicolumn{2}{c|}{\textbf{\dls galaxy bias}} \\
        $\Delta z_{{\rm g}}^3 $  & $\cN (0.0, 0.006)$ &            $b_\mathrm{g}^{i}$   & $U (0.8, 3.0)$ \\
        $\Delta z_{{\rm g}}^4 $  & $\cN (0.0, 0.01)$ &             \multicolumn{2}{c|}{\textbf{\dls photo-z bias}} \\
        $\Delta z_{{\rm g}}^5 $  & $\cN (0.0, 0.01)$  &            $\Delta z_{{\rm g}}^1 $  & $\cN (0.000, 0.007)$ \\
        \multicolumn{2}{|c|}{\textbf{DES source photo-z bias}} &        $\Delta z_{{\rm g}}^2 $  & $\cN (0.000, 0.007)$ \\
        $\Delta z_{{\rm s}}^1 $  & $\cN (-0.001,0.016)$ &          $\Delta z_{{\rm g}}^3 $  & $\cN (0.000, 0.006)$ \\
        $\Delta z_{{\rm s}}^2 $  & $\cN (-0.019,0.013)$ &          $\Delta z_{{\rm g}}^4 $  & $\cN (0.000, 0.01)$ \\\cline{3-4} 
        $\Delta z_{{\rm s}}^3 $  & $\cN (0.009, 0.011)$ &          \multicolumn{2}{c|}{\textbf{KiDS photo-z bias}} \\
        $\Delta z_{{\rm s}}^4 $  & $\cN (-0.018, 0.022)$ &         $\Delta z_{{\rm s}}^1 $  & $\cN (0.000, 0.0106)$ \\ 
        \multicolumn{2}{|c|}{\textbf{DES galaxy bias}}            &    $\Delta z_{{\rm s}}^2 $  & $\cN (0.002, 0.0113)$ \\
        $b_\mathrm{g}^{i}$   & $U (0.8, 3.0)$                 &    $\Delta z_{{\rm s}}^3 $  & $\cN (0.013, 0.0118)$ \\
        \multicolumn{2}{|c|}{\textbf{DES shear calibration bias }}&    $\Delta z_{{\rm s}}^4 $  & $\cN (0.011, 0.0087)$ \\ 
        $m^i $ & $\cN (0.012, 0.023)$                         &    $\Delta z_{{\rm s}}^5 $  & $\cN (-0.006, 0.0097)$\\ 
        \cline{1-2}
        \multicolumn{2}{|c|}{\textbf{Intrinsic Alignments}} &       \multicolumn{2}{c|}{\textbf{KiDS shear calibration bias}}\\
        $A_\mathrm{IA,0} $ & $U(-5.0, 5.0)$&                       $m^1 $  & $\cN (0.000, 0.019)$ \\
        $\eta_\mathrm{IA}$ & $U (-5.0, 5.0)$&                      $m^2 $  & $\cN (0.000, 0.020)$ \\
        $z_0$ & 0.62&                                              $m^3 $  & $\cN (0.000, 0.017)$ \\             
        & &                                                        $m^4 $  & $\cN (0.000, 0.012)$ \\ 
        & &                                                        $m^5 $  & $\cN (0.000, 0.010)$ \\
        \hline
        \end{tabular}
        \caption{Prior distributions for the nuisance parameters entering our analysis for each tracer. $U(a, b)$ and $\cN(\mu, \sigma)$ describe a uniform distribution with boundaries $(a, b)$ and a Gaussian distribution with mean $\mu$ and variance $\sigma$, respectively.  The index $i$ in $b^i_\mathrm{g}$ and $m^i$ runs over the different redshift bins. The DES and intrinsic alignment priors have been taken from Table 1 of  \cite{1810.02322} and the KiDS priors follow \cite{2007.15633}.}\label{tab:priors_nuisance}
      \end{table}

      \begin{itemize}
        \item {\bf Galaxy bias.} We use a linear bias model, assigning a different bias parameter for each galaxy clustering sample and redshift bin (i.e. 5, 4 and 2 parameters for DES$g$, \dls and \eboss respectively).
        \item {\bf Intrinsic alignments.} We use an evolving alignment amplitude (Eq. \ref{eq:ias}), with free amplitude and redshift evolution parameters $A_{\rm IA,0}$, $\eta_{\rm IA}$.
        \item {\bf Photo-$z$s.} We characterize the uncertainties on the redshift distributions of the different galaxy samples in terms of a shift in the mean redshift $\Delta z$. We marginalize over one such parameter in each redshift bin for both clustering and shear samples (except in the case of \eboss), with the same Gaussian prior used by the corresponding collaborations. Note that the analysis of the \dls sample by \cite{2010.00466} used a different parametrization for the redshift distribution uncertainties, stated in terms of the conditional photo-$z$ distribution. The relatively broad priors on $\Delta z$ we use should encompass the small uncertainties found by \cite{2010.00466} after using cross-bin correlations to self-calibrate the redshift distributions, although a more thorough analysis of all possible modes of uncertainty (e.g. in the width of the distribution \cite{1912.08209}) would be desirable both in the case of \des and \dls.
        \item {\bf Quasar shot noise.} As noted in Section \ref{ssec:data.eBOSS}, our estimate of the noise bias for the \eboss sample is not exact. Therefore we marginalize over two parameters characterizing the amplitude of the noise power spectrum for the two quasar redshift bins with a $10\%$ Gaussian prior. Since these parameters are linear in the power spectra, this marginalization can be done analytically by modifying the covariance matrix of the \eboss auto-correlations.
        \item {\bf Multiplicative bias.} We marginalize over a multiplicative bias parameter $m_i$ in each cosmic shear redshift bin, with Gaussian priors derived by \des and \kids. Note that, as done in the \kids analysis \cite{2007.15633}, it is possible to marginalize over these parameters analytically by linearizing the impact of their uncertainty on the power spectra \cite{1606.05338}. We verified that our constraints are not sensitive to the choice of marginalizing over these parameters exactly in the likelihood instead, and thus we do so for consistency with the \des analysis.
      \end{itemize}
      Following the \des Y1 analysis we do not model the impact of baryonic effects on the matter power spectrum. Although this should have a subdominant effect on the scales used here \cite{1809.09148,2007.15632}, a more rigorous study of the cosmological constraints on \lcdm parameters from the combination of all data sets studied here would require a more careful assessment of this source of uncertainty.
      
      Finally, our data vector will contain a combinaion of the following power spectra:
      \begin{itemize}
        \item {\bf Clustering auto-correlations.} We will only consider auto-correlations between galaxies in the same redshift bins. Although the cosmological signal is concentrated in these auto-correlations, cross-bin correlations can be used to self-calibrate photometric uncertainties \cite{1912.08209,2010.00466}. Nevertheless, we choose to discard them in order to mimic the choices made in the \des analysis \cite{1708.01530}. We impose strict scale cuts on clustering, using multipoles smaller than $\ell_{\rm max}=k_{\rm max}/\bar{\chi}$, where $\bar{\chi}$ is the comoving radial distance at the mean redshift of the bin, and $k_{\rm max}=0.15\,{\rm Mpc}^{-1}$. This is done in order to ensure the validity of the linear bias model used here. In addition to this, we will impose large-scale cuts to avoid the impact of large-scale observational systematics in the galaxy auto-correlations. Following \cite{2007.08999,2010.00466}, these correspond to $k_{\rm min}=0.02\,{\rm Mpc}^{-1}$ for \eboss, and $\ell_{\rm min}=30$ for the \dls sample. As done in \cite{1708.01536} we do not include any large-scale cut on the \redmagic sample (and a visual inspection of the power spectra did not reveal any obvious sign of additional large-scale power due to systematics).
        \item {\bf Clustering-lensing cross-correlations.} We will use all available cross-correlations between clustering redshift bins and lensing probes, including both tomographic cosmic shear samples and CMB lensing. The only exception to this is the cross-correlation between \eboss and any of the cosmic shear samples, given their null spatial overlap (see Figure \ref{fig:footprint}). In each cross-spectrum we impose the high-$\ell$ scale cut of the corresponding galaxy clustering sample as described above. 
        \item {\bf Shear-shear correlations.} We use all available cross-correlations between different tomographic bins corresponding to the same cosmic shear sample (ignoring \des-\kids cross-correlations given their zero area overlap). For the \des power spectra we use all bandpowers in the range $30<\ell<2000$, which were shown in \cite{2010.09717} to be free from systematics. For \kids we use a more strict large-scale cut, following \cite{2007.15633,2007.15632}, with $100<\ell<2000$.
        \item {\bf Shear-\cmbk correlations.} We use all cross-correlations between the \planck \cmbk map and the \kids and \des samples, using the same scale cuts used for the analysis of the corresponding shear-shear correlations.
      \end{itemize}
      
      \begin{table}
        \centering
        \def\arraystretch{1.2}
        \begin{tabular}{|ccc|}
        \hline
        \multicolumn{3}{|c|}{\textbf{Scale cuts ($\ell$ and $k$)}} \\
        \hline
        Probe &  min  & max\\  
        \hline 
        \desgc & 0 & $\unit[0.15]{Mpc^{-1}}$\\
        eBOSS QSO & $\unit[0.02]{Mpc^{-1}}$ & $\unit[0.15]{Mpc^{-1}}$\\
            \dls & 30 & $\unit[0.15]{Mpc^{-1}}$\\
        \deswl & 30 & 2000\\
        \kids & 100 & 2000\\
        \cmbk & 8 & 2000\\
        \hline
        \end{tabular}
        \caption{Scale cuts for the different tracers in our data vector. Scales without units are angular $\ell$ modes; whereas those in $\unit[]{Mpc^{-1}}$ are comoving wavenumbers $k$, translated into angular multipoles as $\ell=k\bar{\chi}$, where $\bar{\chi}$ is the comoving distance at the center of each redshift bin. We exclude bandpowers with effective $\ell$s lower or larger than the scale cuts.}\label{tab:scale_cuts}
      \end{table}
      
      The choice of scale cuts used in our analysis is summarized in Table \ref{tab:scale_cuts}. Note that our data vector does not include the auto-correlation of the \planck \cmbk map. Although this would provide a valuable constraint on the integrated evolution of cosmic structures over the range of redshifts studied, we choose to exclude it in order to avoid the complications of modelling the different cosmology-dependent lensing biases \cite{1611.09753,1807.06210} as well as accurately describing the covariance of this power spectrum with all other components of our data vector, given the spatial overlap of the lensing map with all our data sets.
      
\begin{figure}
        \centering
        \includegraphics[width=\textwidth]{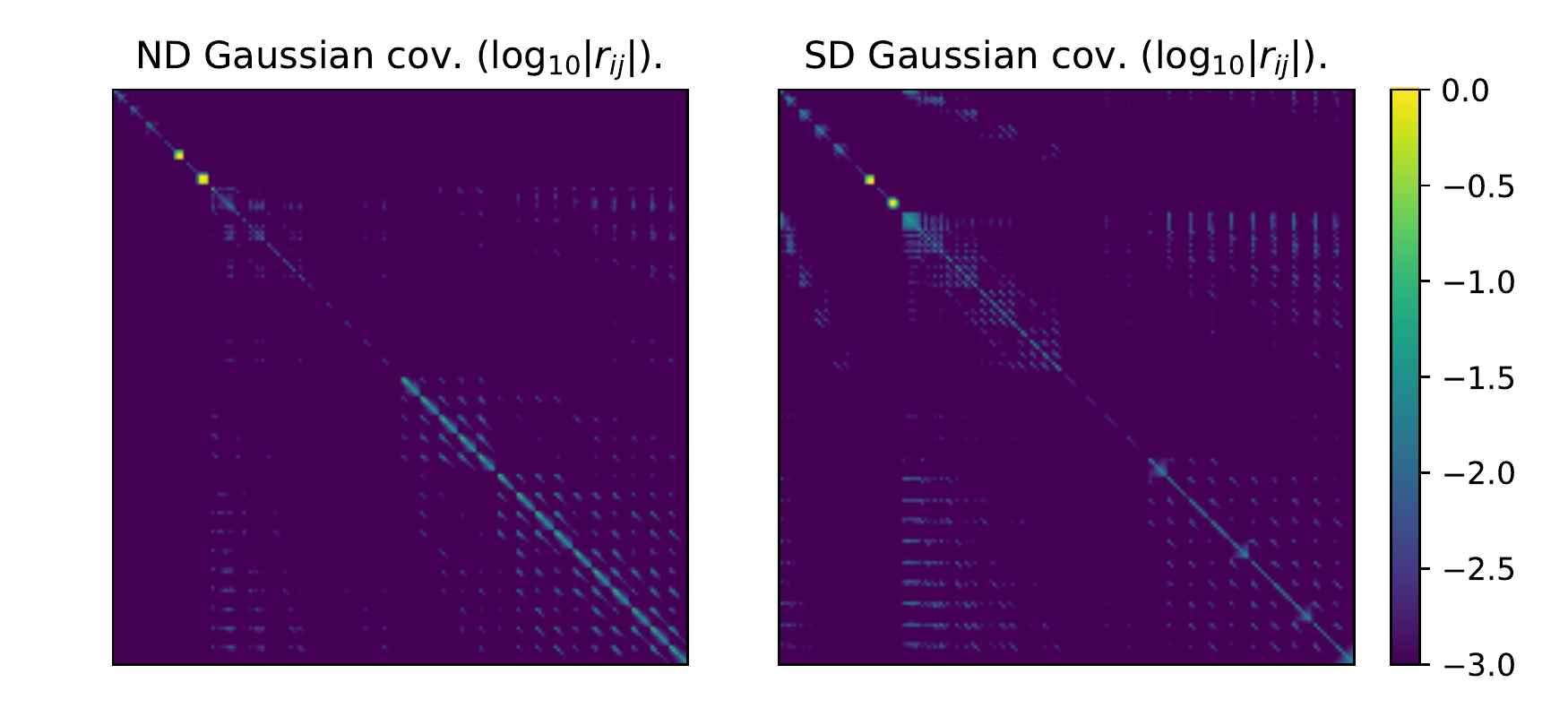}
        \caption{Correlation matrices ($r_{ij}$) of the \northd (left) and \southd (right) power spectra once the scale cuts in Table~\ref{tab:scale_cuts} have been applied as they enter into the MCMC. The yellow points correspond to \eboss auto-correlations, and are caused by the analytical noise marginalization. We see that the Gaussian covariance is mainly described by its diagonal elements, although there are some non-negligible correlations between nearby elements.}\label{fig:covG_corr}
      \end{figure}

      The resulting data vector contains 665 and 662 elements for the \southd and \northd data sets respectively, and 1275 elements in the case of the full combination. Figure~\ref{fig:covG_corr} shows the correlation matrices for the \northd and \southd data sets from their Gaussian covariance matrices after applying the scale cuts.
      
  \section{Results}\label{sec:res}
    \subsection{Validation}\label{ssec:res.val}
      Since our constraints will be based on a re-analysis of the different data sets presented in Section \ref{sec:data} using a common harmonic-space framework, we carry out a set of basic validation tests to ensure the robustness of the resulting power spectra and covariance matrices. 
  
      \nmt{}, the pseudo-$C_\ell$ power spectrum estimator we use, has been extensively validated \cite{1809.09603,1906.11765}, and the specific case of cosmic shear data, given the significantly more complex survey geometry, was studied in detail in \cite{2010.09717}. The main object of our validation is therefore to diagnose the presence of systematics in the maps that could affect the estimated power spectra.

      Since the cosmological shear signal is dominated by a pure $E$ mode (although see e.g. \citep{0904.4703} ), we study the significance of any power spectra involving shear $B$-modes as a signature of systematic contamination. We do so by calculating, for each such power spectrum, the probability-to-exceed ($p$-value) of its $\chi^2$ with respect to a null signal. The results are shown in Table \ref{tab:null_tests} for all the combinations involving shear tracers explored here. In the vast majority of cases the resulting $p$-values are acceptable, showing no evidence of $B$-mode contamination at more than 2$\sigma$ ($p>0.05$). Although we find 4 cases with $p<0.05$, such a small number is compatible with the look-elsewhere effect. We quantify this by conducting a Kolmogorov-Smirnov test on the full set of $\chi^2$ values with respect to a $\chi^2$ distribution, finding probabilities $p=0.72$ and $p=0.35$ for the \northd and \southd data sets respectively. Therefore we find no evidence of $B$-modes in either of the shear samples within the range of scales used here.
      
        \begin{table}[tb]
            \small
            \centering
            \begin{tabular}{|l|ccccc||l|cccc|}
                \hline
                Tracer   &\multicolumn{5}{c||}{\kids} &
                Tracer   &\multicolumn{4}{c|}{\des $\shear$}\\
                name     & Bin 0 & Bin 1 & Bin 2 & Bin 3 & Bin 4 &
                name     & Bin 0 & Bin 1 & Bin 2 & Bin 3 \\
                \hline
                \dls-0   & 0.460 & 0.135 & 0.234 & 0.978 & 0.650 & \des$g$-0 & 0.396 & 0.733 & 0.704 & 0.294 \\
                \dls-1   & 0.011 & 0.781 & 0.661 & 0.105 & 0.438 & \des$g$-1 & 0.737 & 0.983 & 0.889 & 0.071 \\
                \dls-2   & 0.226 & 0.425 & 0.752 & 0.163 & 0.861 & \des$g$-2 & 0.378 & 0.809 & 0.264 & 0.288 \\
                \dls-3   & 0.483 & 0.324 & 0.567 & 0.569 & 0.269 & \des$g$-3 & 0.923 & 0.073 & 0.905 & 0.354 \\
                \cmbk    & 0.280 & 0.050 & 0.078 & 0.167 & 0.450 & \des$g$-4 & 0.517 & 0.048 & 0.889 & 0.459 \\
                \kids-0  & 0.949 & 0.604 & 0.463 & 0.586 & 0.761 & \cmbk     & 0.168 & 0.170 & 0.432 & 0.943 \\
                \kids-1  &  --   & 0.795 & 0.292 & 0.877 & 0.336 & \des$\gamma$-0 & 0.436 & 0.232 & 0.630 & 0.774 \\
                \kids-2  &  --   &  --   & 0.603 & 0.044 & 0.006 & \des$\gamma$-1 &  --   & 0.545 & 0.991 & 0.645 \\
                \kids-3  &  --   &  --   &  --   & 0.977 & 0.406 & \des$\gamma$-2 &  --   &  --   & 0.813 & 0.245 \\
                \kids-4  &  --   &  --   &  --   &  --   & 0.612 & \des$\gamma$-3 &  --   &  --   &  --   & 0.977 \\
                \hline
            \end{tabular}
            \caption{Null tests for all power spectra involving shear $B$ modes. Each cell displays the $p$-value for the $\chi^2$ of the corresponding power spectrum with respect to the null hypothesis. Results are shown for the \northd and \southd data combinations in the left and right halves of the table respectively. The entries involving two shear samples contain the $p$-value for a data vector combining all possible $B$-$B$ and $E$-$B$ power spectra.}
            \label{tab:null_tests}
        \end{table}
  
      Another source of map-level systematics is the modification of the observed number density of galaxies due to various observational systematics (e.g. dust absorption, star contamination, completeness variations). The impact of these systematics has, to some extent, already been taken into account in the three galaxy clustering samples used in this analysis. The \des \redmagic galaxies and \eboss sample have galaxy weights associated to them aimed at correcting for the effect of known systematics, and the overdensity maps for the \dls sample used here are corrected by completeness variations as well as a fifth-order polynomial of the local star density \cite{2010.00466}. Additionally, we implement large-scale cuts on the auto-correlations of the \eboss and \dls data sets following the prescription presented in \cite{2007.08999} and \cite{2010.00466} respectively, removing scales significantly affected by these systematics (see Table \ref{tab:scale_cuts}). In order to test for the impact of residual contamination, we recompute these power spectra making use of the linear deprojection method implemented in \nmt. The method removes the impact of specific systematics by projecting the data on the subspace orthogonal to a set of known contaminant maps, correcting for the resulting bias to the power spectrum analytically (see \cite{1609.03577,1809.09603}). We then compare the resulting power spectra with those calculated without deprojection and calculate their relative $\chi^2$ in order to detect significant deviations. For \eboss we deproject a set of 15 systematic templates corresponding to observing conditions in the SDSS survey as well as a dust template based on \cite{astro-ph/9710327}. For \dls, we deproject the completeness and star density maps made publicly available by \cite{2010.00466}. For the \redmagic sample, we deproject the set of 13 most relevant survey property maps identified in \cite{1708.01536}. In all cases we do not observe any significant deviation within the range of scales used here, with a relative reduced $\chi^2$ below 0.1 in all cases.

      The methods used to estimate the Gaussian covariance matrix have been thoroughly validated in \cite{1906.11765,2010.09717} for the type of data used here. This, together with the acceptable $\chi^2$ values we find in Section \ref{ssec:res.lcdm}, as well as a visual inspection of the scatter in the power spectrum residuals with respect to the \lcdm best-fit prediction (see Section \ref{ssec:res.lcdm}) reassure us that the covariance matrix used here is not significantly over- or under-estimated.

    \subsection{\lcdm constraints}\label{ssec:res.lcdm}

      In this Section we discuss how these data sets constrain the \lcdm model. On the one hand, this is a validating exercise through which we show that our results are in agreement with those found by others in the literature using different combinations of our data. On the other hand, it also allows us to present \lcdm constraints from a novel combination of data sets.

      \begin{figure}
        \centering
        \includegraphics[width=0.49\textwidth]{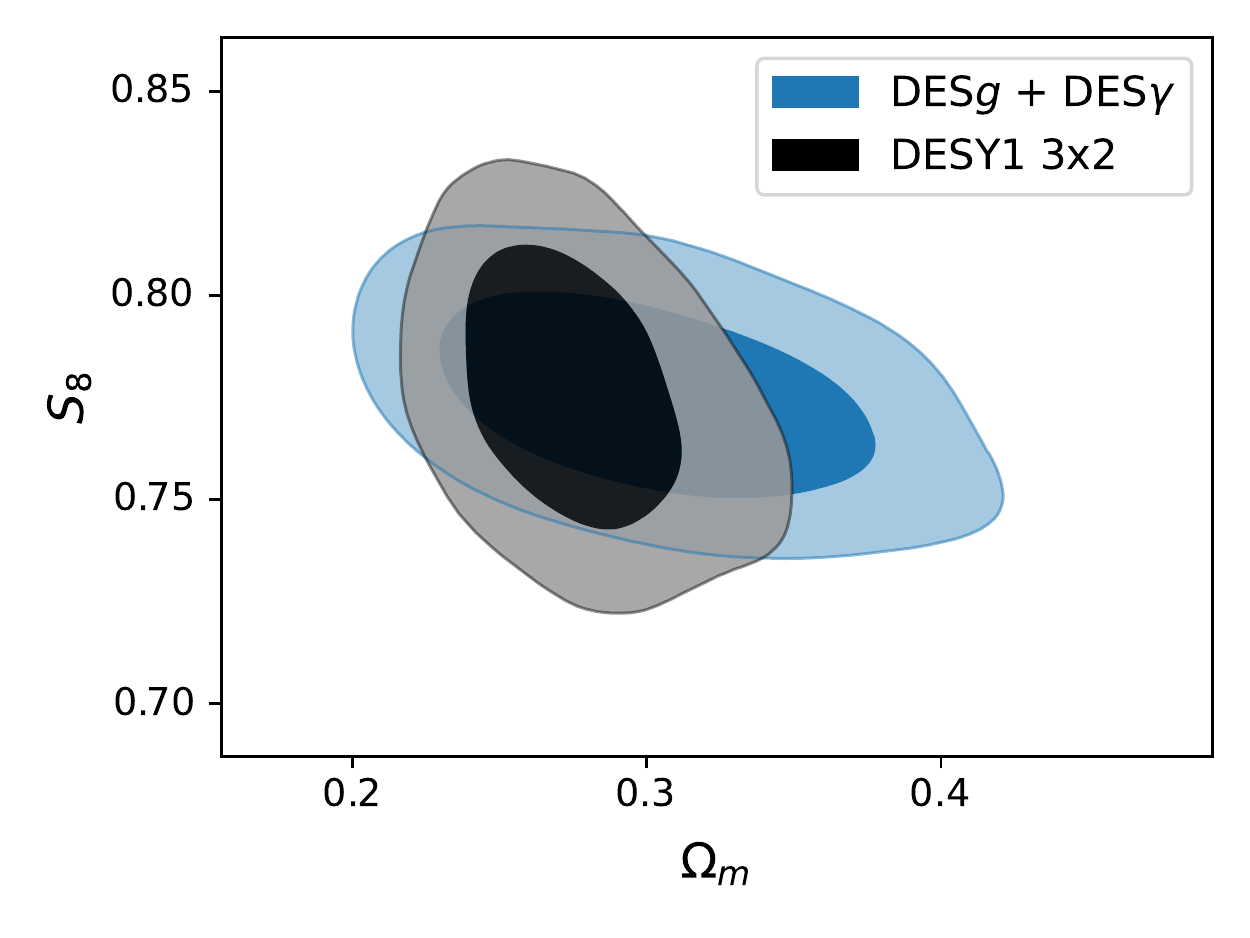}
        \includegraphics[width=0.49\textwidth]{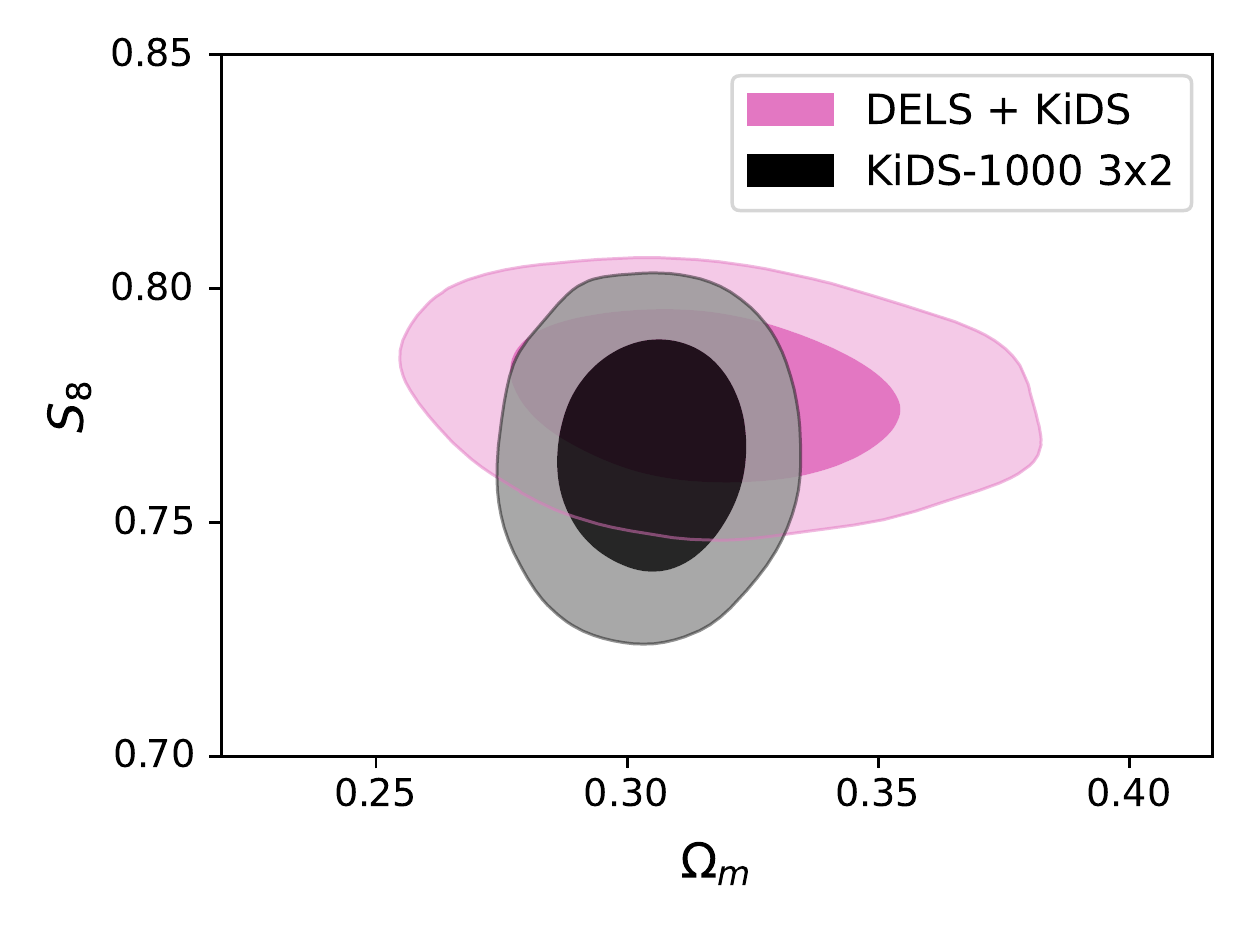}
        \caption{$S_8$-$\Om$ posterior distributions at 68 and 95\% Confidence Level (C.L.) from our $3\times2$ analysis of the \des data (left, blue) and of the \kids+\dls data (right, pink). In both cases, we compare our constraints with those found by the respective collaborations~\cite{1708.01530, 2007.15632} using different analysis choices (and different galaxy clustering data in the case of \kids).}
        \label{fig:ND_SD_lcdm_s8}
      \end{figure}

      Figure \ref{fig:ND_SD_lcdm_s8} shows the constraints on $\Om$ and $S_8$ for the $3\times2$pt analysis of the \des and \kids+\dls samples (blue and pink contours, respectively), together with the results found in the official $3\times2$pt analysis published by the \des and \kids collaborations. We find an overall good agreement between both sets of constraints, although our analysis recovers visibly larger uncertainties in both cases. These differences are understandable, since the analyses are not equivalent. The official \des analysis~\cite{1708.01530} made use of real-space correlation functions, making it difficult to match their choice of scale cuts. The \kids $3\times2$pt analysis~\cite{2007.15632} made use of spectroscopic clustering samples (BOSS and 2dFLenS), as well as a variety of two-point function estimators. Furthermore, unlike these official analyses, we considered only massless neutrinos and did not marginalize over their mass. The poorer constraints we find on $\Om$ in comparison with the \kids analysis are expected, since our use of photometric clustering prevents us from taking advantage of the BAOs as a standard ruler. As shown in \cite{2103.09820} this could be improved significantly by increasing the range of scales over which galaxy clustering can be used through a perturbative bias expansion. The method used here to estimate the power spectrum covariance matrix, as well as the choice of parameter priors, also differ from those used by both collaborations. Nevertheless, the comparison with these published results for the \lcdm model, together with the validation tests described in the previous section, does not reveal any significant issues in our analysis.

      \begin{figure}
        \centering
        \includegraphics[width=0.49\textwidth]{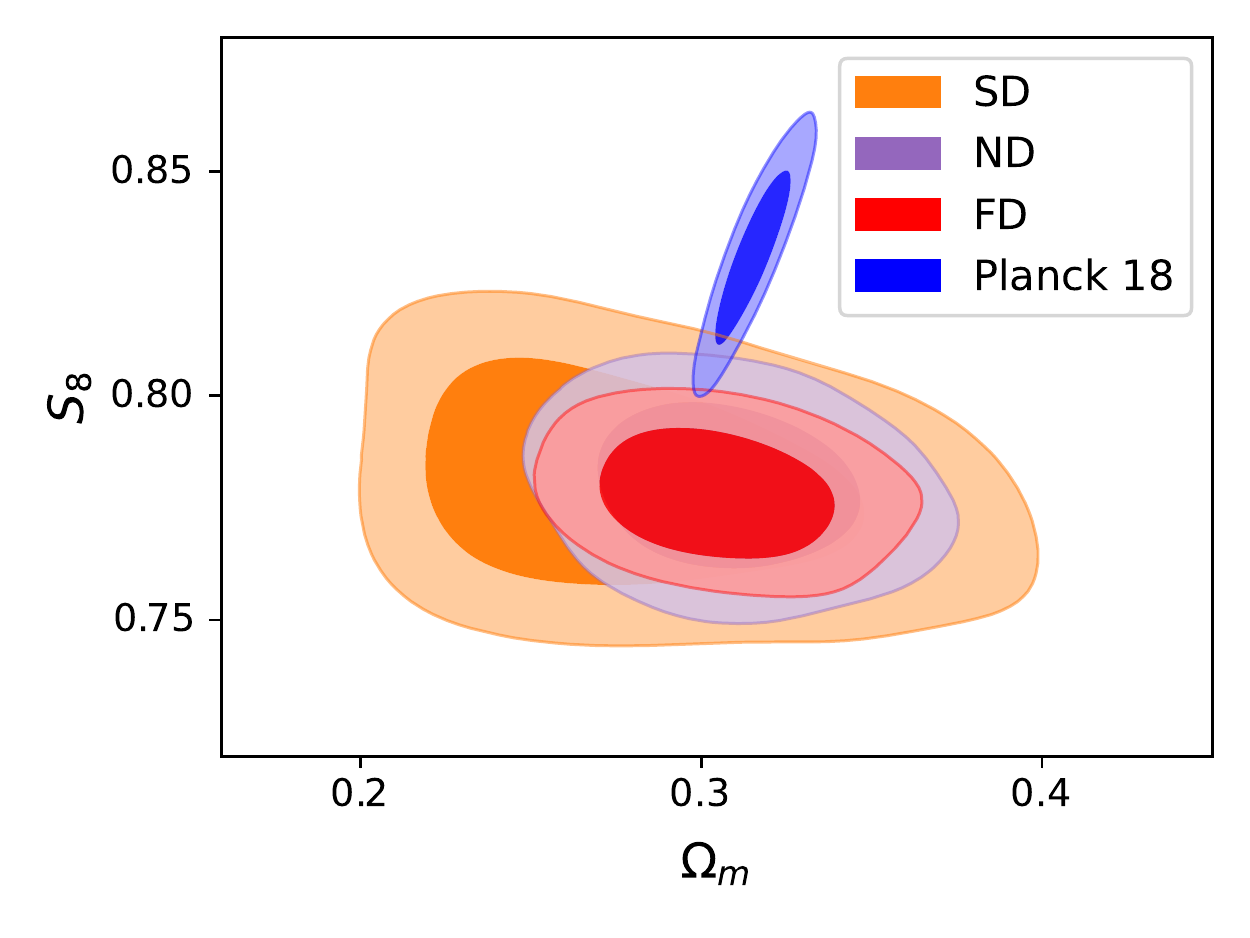}
        \includegraphics[width=0.49\textwidth]{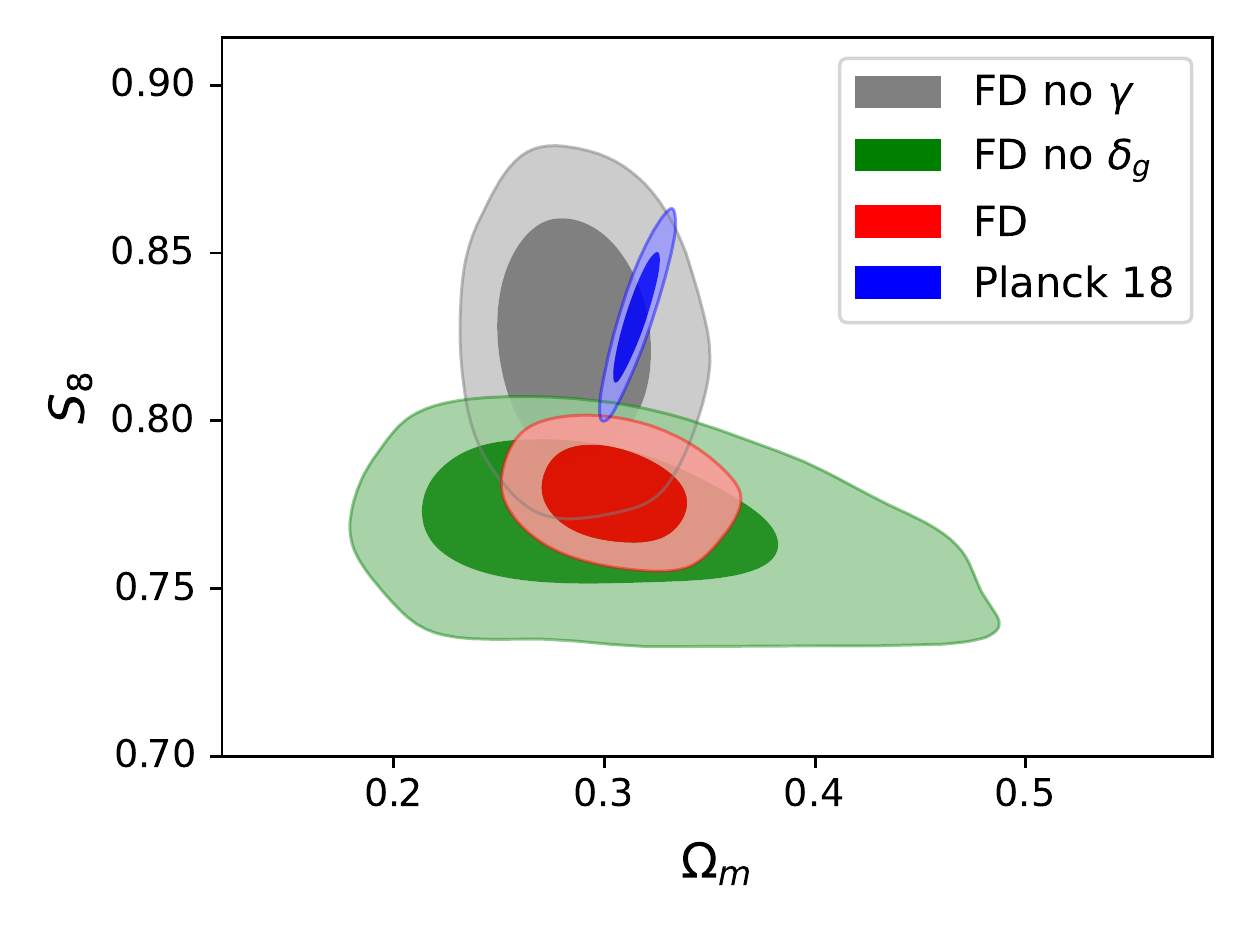}
        \caption{{\sl Left:} $S_8$-$\Om$ posterior distributions at 68 and 95\% C.L. for the \northd, \southd and \alld data sets together with the \planck CMB constraints (purple, orange, red and blue respectively). {\sl Right:} constraints from our full data set (red), excluding galaxy clustering (green), and excluding cosmic shear (gray), in addition to the \planck constraints (blue). Here, \southd and \northd stand for \des + \eboss + \cmbk and \dls + \kids + \eboss + \cmbk, respectively; while \alld is the combination of all datasets.}
        \label{fig:all_lcdm_s8}
      \end{figure}
 
\begin{figure}
          \centering
          \includegraphics[width=\textwidth]{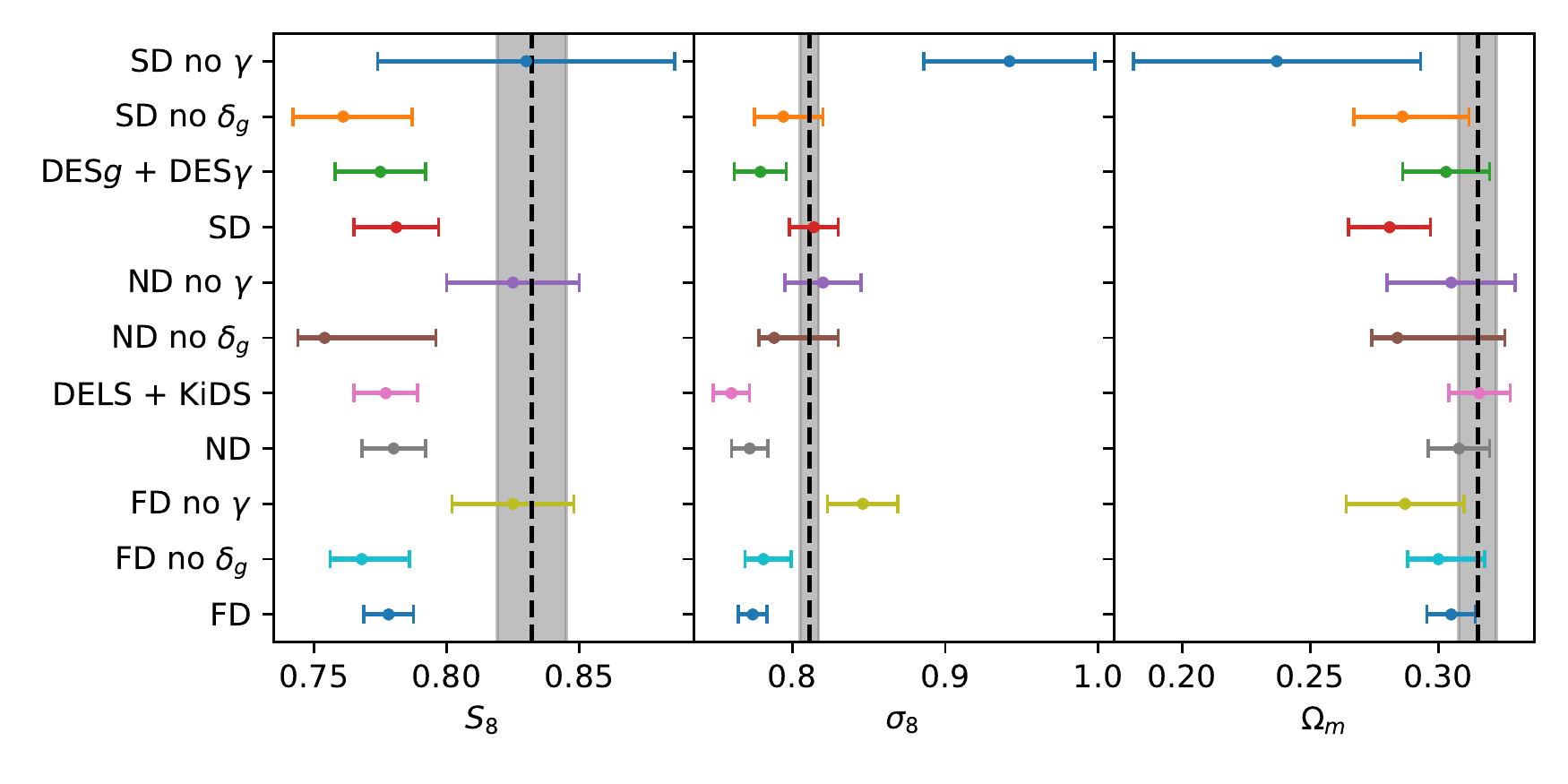}
          \caption{Main results from the \lcdm constrains. Marginalized posterior distributions with 68\% C.L. error bars for the $S_8$, $\sigma_8$ and $\Om$ parameters. The black dashed line and gray area correspond to the \planck 18 values of these parameters. See Table~\ref{tab:chi2} for a quantitative representation of these results. }
          \label{fig:lcdm_results}
      \end{figure}

      The left panel of Figure \ref{fig:all_lcdm_s8} shows the final constraints found from different combinations of the full data set in the $(\Om, S_8)$ plane. Results are shown for the combination of \northd (purple), \southd (orange) and \alld (red) experiments. The \planck \lcdm constraints are shown in blue. The right panel of Figure \ref{fig:all_lcdm_s8} then breaks these constraints down by tracer combination, showing the constraints found in the absence of shear (gray) and galaxy clustering (green) in addition to the full data set (red) and \planck (blue). These constraints on $S_8$ and $\Om$ are also listed in Table~\ref{tab:chi2} for the different experiment combinations explored here, as well as pictorially represented in Figure~\ref{fig:lcdm_results}. Although in most cases we find an increasing tension with the value of $S_8$, it is interesting to note that this tension is driven by the shear data, and is not evident through the combination of galaxy clustering and CMB lensing. Note, however, that this tension has also been reported from this probe combination (although using different data sets) by other groups \cite{2010.00466,2105.03421}. In spite of this tension, the different data sets used here are in reasonable agreement with each other. Since there is no obvious sign of tension between them, we combine them to find a constraint on $S_8$ given by
      \begin{equation}
        S_8=0.7781\pm0.0094\hspace{12pt}(68\%\,\,{\rm C.L.}).
      \end{equation}
      Compared with the constraints found by \planck on this parameter, $S_8^\text{\planck}=0.832\pm0.013$, and assuming Gaussian errors added in quadrature, the level of tension is $\sim3.4\sigma$. This is 0.4$\sigma$ larger than the tension found by the \kids collaboration \cite{2007.15632,2105.09545}, and in agreement with previous results. As noted in \cite{1902.04029,2007.15632} tension between experiments in their measurement of one particular parameter is not necessarily indicative of tension between their data sets when the full multi-dimensional parameter space is taken into account. Nevertheless, since similar constraints have been consistently obtained by various groups, using different data sets, this parameter tension must be analyzed further. Although more insight is expected from the ongoing analysis of new data from \des and other collaborations, this motivates our reconstruction of the growth history as a means to better understand the origin of this tension.

     \begin{table}
        \centering
        \def\arraystretch{1.2}
        \begin{tabular}{|lcccccc|}
        \hline
        Probes        &  $N_d$  & $\chi^2$ & $p$  & $S_8$  & $\sigma_8$ & $\Om$\\
        \hline 
        \southd No $\shear$ & 141 & 111 & 90 & $ 0.830\pm 0.056$ & $ 0.942\pm 0.092$ & $ 0.237^{+0.019}_{-0.040}$ \\
        \southd No $\delta_g$ & 336 & 347 & 30 & $ 0.761^{+0.026}_{-0.019}$ & $ 0.794\pm 0.091$ & $ 0.286^{+0.042}_{-0.077}$ \\
        \des & 475 & 496 & 18 & $ 0.775\pm 0.017$ & $ 0.779^{+0.063}_{-0.080}$ & $ 0.303^{+0.044}_{-0.053}$ \\
        \southd  & 665 & 660 & 45 & $ 0.781\pm 0.016$ & $ 0.814^{+0.074}_{-0.065}$ & $ 0.281^{+0.029}_{-0.054}$\\
        \hline
        \northd No $\shear$ & 112 & 91 & 81 & $ 0.825\pm 0.025$ & $ 0.820\pm 0.044$ & $ 0.305^{+0.023}_{-0.028}$ \\
        \northd No $\delta_g$ & 440 & 460 & 23 & $ 0.754^{+0.042}_{-0.010}$ & $ 0.788^{+0.095}_{-0.078}$ & $ 0.284^{+0.046}_{-0.079}$ \\
        \dls + \kids & 610 & 649 & 10 & $ 0.777\pm 0.012$ & $ 0.760\pm 0.035$ & $ 0.316^{+0.023}_{-0.027}$ \\
        \northd & 662 & 688 & 18 & $ 0.780\pm 0.012$ & $ 0.772\pm 0.036$ & $ 0.308^{+0.023}_{-0.027}$ \\
        \hline
        \alld No $\shear$ & 201 & 177 & 71 & $ 0.825\pm 0.023$ & $ 0.846\pm 0.045$ & $ 0.287^{+0.022}_{-0.027}$ \\
        \alld No $\delta_g$ & 776 & 806 & 20 & $ 0.768^{+0.018}_{-0.012}$ & $ 0.781^{+0.086}_{-0.076}$ & $ 0.300^{+0.041}_{-0.071}$ \\
        \alld & 1275 & 1312 & 16 & $ 0.7781\pm 0.0094$ & $ 0.774\pm 0.033$ & $ 0.305^{+0.021}_{-0.025}$ \\
        \hline
        \end{tabular}
        \caption{Main results from the \lcdm constraints. All results are shown with 68\% C.L. errors. The effective number of degrees of freedom is estimated as $N_{\rm dof,eff}=N_d-N_b-2$, where $N_d$ is the total number of data points, and $N_b$ is the number of free galaxy bias parameters. See Figure~\ref{fig:lcdm_results} for a visual representation of these results.}\label{tab:chi2}
      \end{table}
      
\begin{figure}
        \centering
        \includegraphics[width=0.6\textwidth]{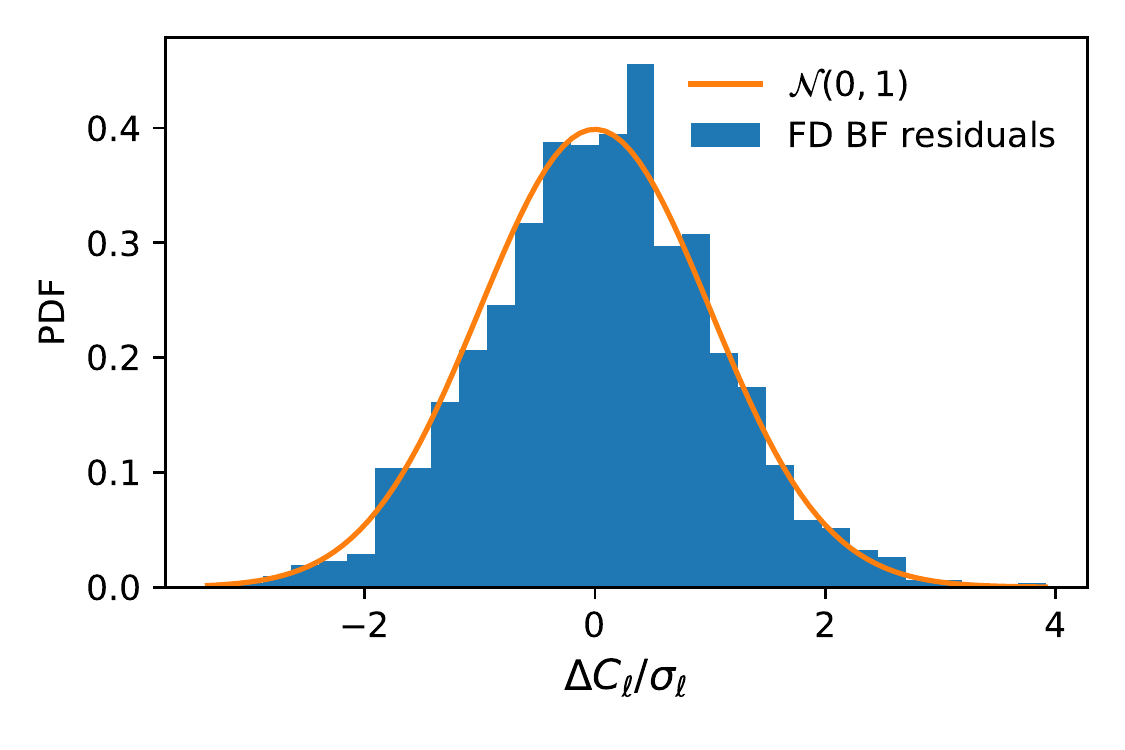}
        \caption{Distribution of the \alld power spectrum residuals with respect to the best-fit \lcdm model normalized by their standard deviation. The distribution agrees well ($p = 0.04$ in the Komogorov-Smirnov test) with a standard normal distribution with zero mean and unit variance.}\label{fig:res_normal}
      \end{figure}

      Table~\ref{tab:chi2} also lists the values of the $\chi^2$ as a metric for goodness of fit for each data set combination. In each case we calculate the corresponding $p$-value assuming an effective number of degrees of freedom given by $N_{\rm dof,eff}=N_d-N_b-2$, where $N_d$ is the total number of data points, and $N_b$ is the number of free galaxy bias parameters. This was found to be a reasonable rule of thumb in \cite{2103.09820}, and we expect it to be still the case since only two cosmological parameters, $\Om$ and $S_8$, are completely constrained by the data.  In all cases we find reasonable $p$-values above 10\%. The \lcdm model therefore is able to describe the data reasonably well, and we find no signatures of strong tension between data sets. Note that achieving this satisfactory goodness of fit is remarkable given the difficulties that have traditionally been involved in the calculation of the covariance matrix for two-point correlations \citep{1708.01530,1804.10663,2007.15633,2012.08568}. This is likely the result of a careful accounting for the impact of survey geometry on all signal and noise contributions to the covariance matrix combined with the relatively simpler covariance calculation needed in a harmonic-space pipeline. In order to ensure that this is not the result of an over-estimation of the power spectrum uncertainties, we visually inspect all the power spectrum residuals with respect to the best-fit \lcdm model using all data, revealing no obvious trend or signs of error overestimation. More quantitatively, a Komogorov-Smirnov test reveals that the distribution of the residuals normalized by their estimated standard deviation agrees well ($p = 0.04$) with a standard normal distribution with zero mean and unit variance (see also Figure \ref{fig:res_normal}). Thus we find no evidence of misestimation in the power spectrum uncertainties.

    \subsection{Growth reconstruction}\label{ssec:res.growth}
     \begin{figure}
        \centering
        \includegraphics[width=0.9\textwidth]{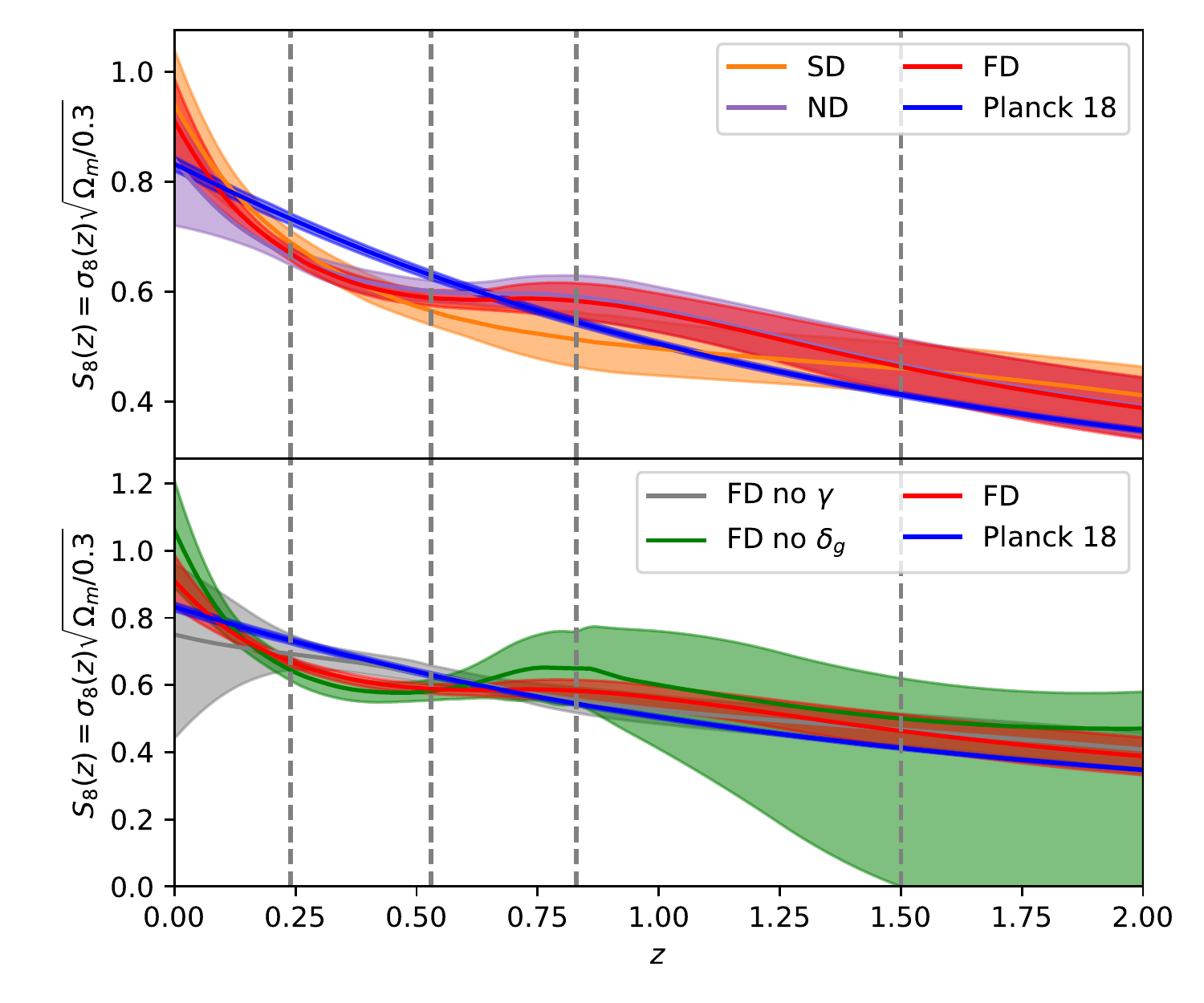}
        \caption{Reconstructed growth history. Each band shows the 68\% C.L. constraints from different data combinations, obtained interpolating with a quadratic spline the upper and lower $1 \sigma$ constraints for different values of $S_8(z_i)$ at a fine grid of redshifts $z_i$. These were computed at each step of the MCMC as derived parameters and their posterior distributions obtained through the same pipeline as the free parameters using \texttt{GetDist}. The top panel shows the constraints from the \northd and \southd data sets (purple and orange respectively), as well as the full combination \alld (red), and the \lcdm constraints from \planck (blue). The bottom panel additionally shows results for the \alld data excluding galaxy clustering (green) or cosmic shear (gray). The vertical dashed lines show the position of the redshift nodes used here to generated the growth factor spline. The $S_8$ tension can be seen at $z\sim 0.4$.}\label{fig:all_s8z}
      \end{figure}
      
      \begin{figure}
        \centering
        \includegraphics[width=0.6\textwidth]{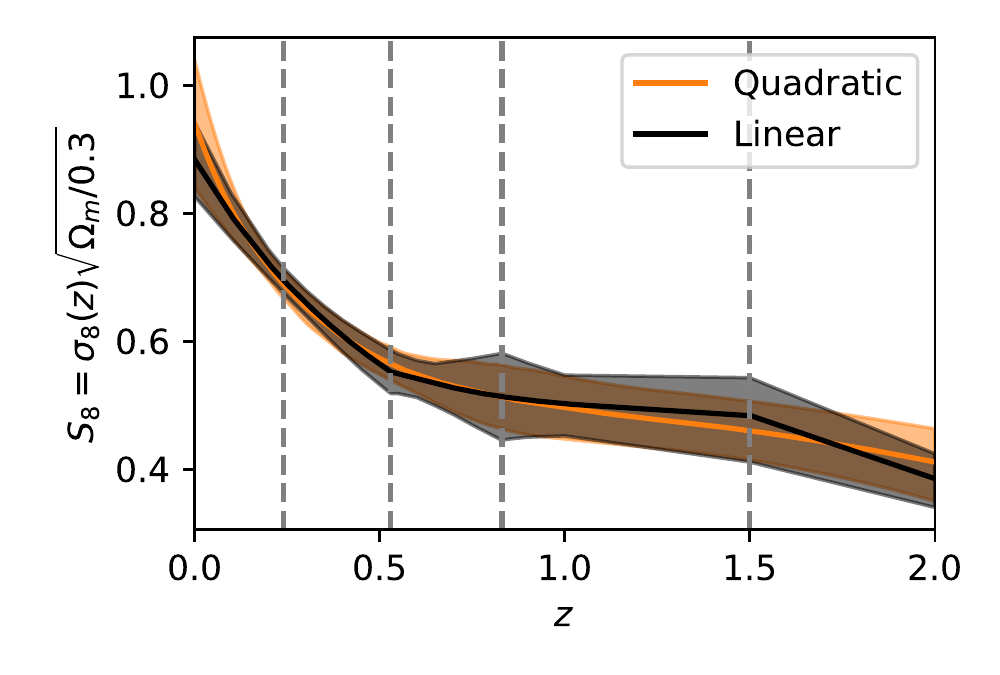}
        \caption{Reconstructed growth history from the \southd data using our fiducial quadratic spline (orange) and a linear spline (gray) showing 68\% C.L. errors. Both methods recover compatible growth histories with slight variations that become most prominent at low redshifts beyond the interpolation range.}\label{fig:splines}
      \end{figure}

      We now move on to the constraints on the linear growth of structure found from the different data combinations explored here. As a reminder, we model the redshift-dependence of the linear growth factor in terms of its value at 4 redshift nodes, $z = \{0.24, 0.53, 0.83, 1.5\}$, from which we interpolate and extrapolate to other redshifts using a quadratic spline (see Section \ref{ssec:theory.growth_rec} for further details).

      The top panel of Figure~\ref{fig:all_s8z} shows the amplitude of matter fluctuations as a function of redshift, parametrized by $S_8(z)$ defined as
      \begin{equation}
        S_8(z)\equiv\sigma_8(z)\sqrt{\Om/0.3},
      \end{equation}
      where $\sigma_8(z)=D(z)\,\sigma_8^{\rm fid}$, $\sigma_8^{\rm fid}=0.8111$ and $\Om$ is the present value of the matter density parameter. This allows us to visualize the evolution of structure growth as directly measured by the data while setting the amplitude to the parameter that these data sets are able to constrain best. Results are shown for the \northd (purple) and \southd (orange) data sets, as well as the full \alld combination (red), together with the growth history predicted by \planck in the \lcdm model (blue). The bottom panel of the same figure shows the same reconstructed growth for different probe combinations, excluding shear data (gray) and galaxy clustering (green). As the figure shows, most of the constraining power from these projected data sets comes from the redshift range $0.25\lesssim z\lesssim 0.7$, through the combination of photometric clustering and cosmic shear, and, to a lesser extent, at $z\sim2$ from the combination of QSO clustering and CMB lensing. By comparison with the \planck \lcdm prediction, we see that the tension between current large-scale structure and CMB data originates around $z\sim0.4$, where the data display a significantly lower amplitude of fluctuations. Furthermore, this feature is recovered consistently by both the \northd and \southd samples independently. Furthermore, this tension seems to be driven by the cosmic shear data, particularly \kids, whereas the combination of galaxy clustering and CMB lensing seems to be in reasonable agreement with \planck (although see \cite{2105.03421}). The current uncertainties, however, are too large to show any tension between these two probes.
      
      The complementarity between weak lensing and galaxy clustering data allows us to reconstruct the growth history from $z\sim2$. Nevertheless, more data at low ($z<0.2$) and high ($z>1$) redshifts would be necessary in order to fully recover the evolution of matter fluctuations, and better understand the origin of the slower growth potentially preferred by current data at intermediate redshifts.

      When interpreting the reconstructed growth history it is important to bear in mind the potential impact of the choice of interpolation scheme (quadratic spline) on the results. This could have an effect both on the recovered growth parameters and on the extrapolation of the growth factor in redshift ranges where we do not have direct measurements. To quantify this, we have repeated our analysis of the \southd data set using linear interpolation between nodes. The result, shown in Figure~\ref{fig:splines}, demonstrates that the choice of interpolation scheme does not have a significant effect on the recovered growth constraints in the range where the galaxy clustering and weak lensing data lie. However, the amplitude of perturbations predicted at $z\sim0$, as well as its statistical uncertainties can be significantly affected by this choice, and therefore care should be taken when interpreting these constraints beyond the range of redshifts covered by the data. In addition, we have checked that we recover the same $\chi^2$ for the best fit \lcdm \southd, \northd and \alld cases if we compute it with the reconstructed growth model using the $\tilde{D}(z_i)$ values found in \lcdm as the spline nodes. More precisely, the relative deviations of the $\chi^2$ between both cases for the \southd, \northd and \alld datasets are -0.06\%, -0.02\%, -0.005\%, respectively. Finally, we also checked that we are able to recover the fiducial cosmology when using simulated data in our MCMC.
      \begin{figure}
        \centering
        \includegraphics[width=\textwidth]{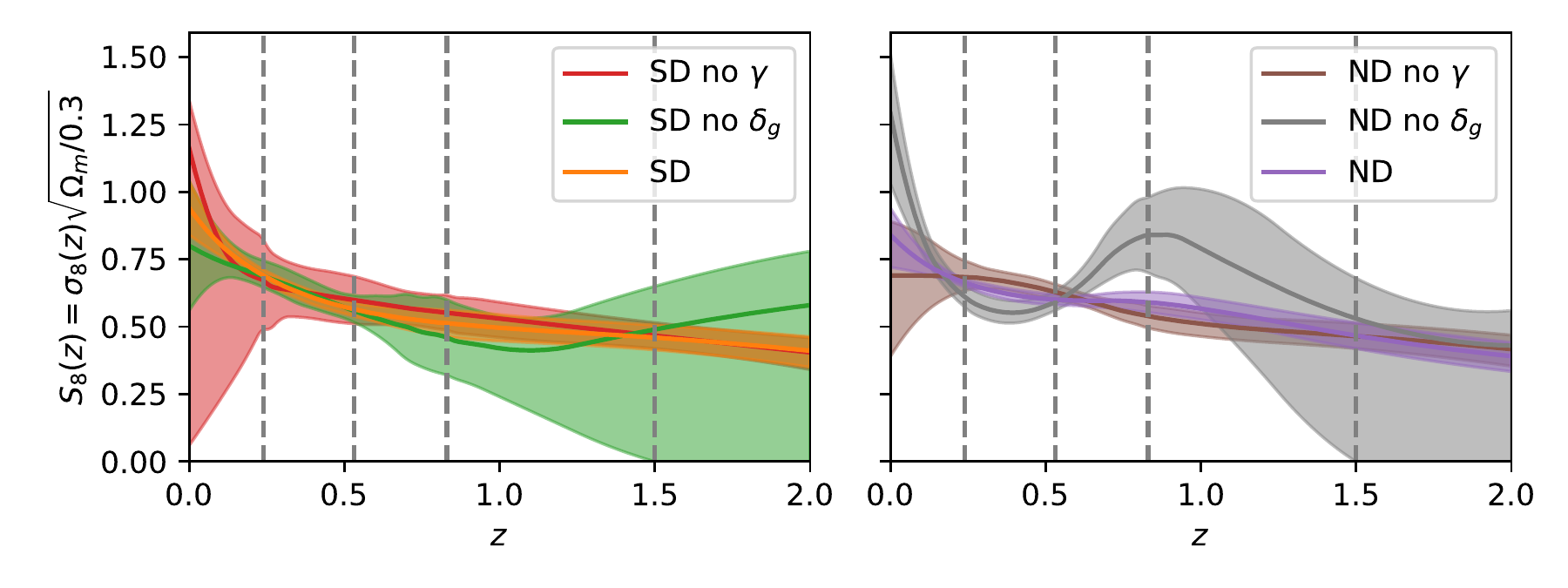}
        \caption{Recovered 68\% CL constraints on the evolution of the linear density fluctuations. The left and right panels show results from the \southd and \northd data sets. The full combination is shown in orange and purple respectively, while the results excluding galaxy clustering and cosmic shear are shown in green/gray and red/brown respectively. The vertical dashed lines show the position of the redshift nodes used here to generated the growth factor spline.}
        \label{fig:ND_SD_s8z}
      \end{figure}
      \begin{figure}
        \centering
        \includegraphics[width=0.9\textwidth]{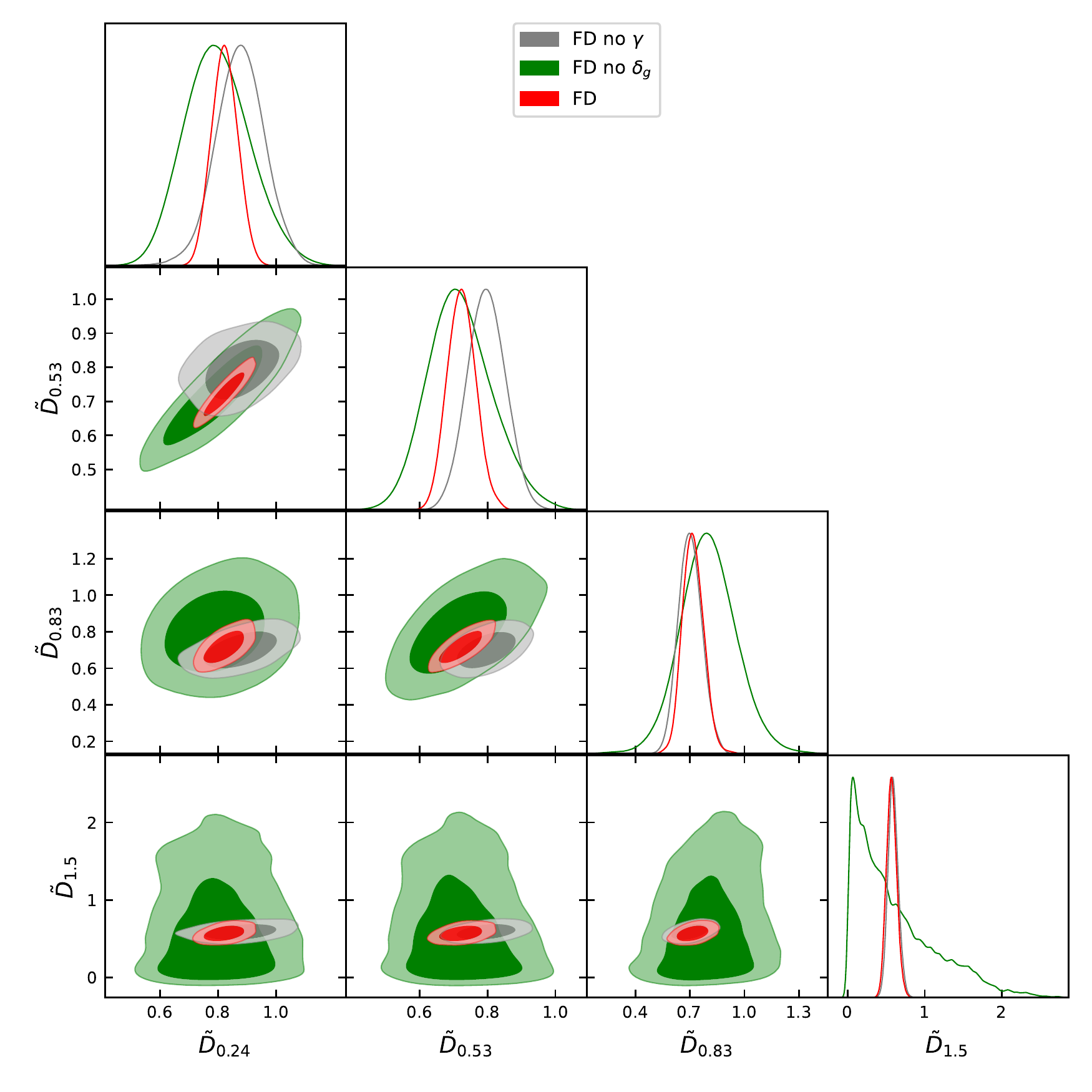}
        \caption{Posterior contours at 68 and 95\% C.L. for the reconstructed growth nodes. Results are shown for different data combinations of the \alld data set. The full constraints are shown in orange, while the green and red contours show the constraints in the absence of galaxy clustering or cosmic shear respectively. Due to its cumulative nature, constraints from cosmic shear produce a visible correlation between adjacent redshift nodes. The constraints on the highest-redshift node are completely dominated by the combination of \eboss and CMB lensing.}\label{fig:DES_dpk}
      \end{figure}

      It is interesting to understand in better detail the redshift ranges over which the different data sets studied here constrain structure growth. To this end, the left and right panels of Figure~\ref{fig:ND_SD_s8z} show the constraints on $S_8(z)$ from different combinations of the \northd and \southd experiments respectively. As the figure shows, the constraints at low redshifts ($z<0.5$) are dominated by the cosmic shear data sets, while the intermediate redshift regime $0.5\lesssim z\lesssim2$ is dominated by the combination of galaxy clustering and CMB lensing. Both data sets are thus highly complementary and their combination is able to significantly improve the accuracy of the reconstruction over the full redshift range. This is understandable given the cumulative nature of weak lensing, which weights it more heavily towards low redshifts. At high redshifts, the overlap between the galaxy clustering and shear kernels is reduced, whereas the overlap with the CMB lensing kernel is equally significant at all redshifts. The figure also shows that the slight kink at $z\simeq0.8$ visible in Figure~\ref{fig:all_s8z} originates from the \kids cosmic shear data, likely due to a statistical fluctuation.
        \begin{table}[tb]
            \centering
            \begin{tabular}{|lllll|}
                \hline
                Probes        & $\tilde{D}_{0.24}$ &  $\tilde{D}_{0.53}$ &  $\tilde{D}_{0.83}$ &  $\tilde{D}_{1.5}$\\
                \hline 
                \southd No $\shear$ & $ 0.92^{+0.21}_{-0.27}$ & $ 0.83\pm 0.15$ & $ 0.76^{+0.11}_{-0.12}$ & $ 0.646\pm 0.085$ \\
                \southd No $\delta_g$ & $ 0.96^{+0.15}_{-0.18}$ & $ 0.78^{+0.12}_{-0.17}$ & $ 0.64^{+0.19}_{-0.23}$ & $ 0.66^{+0.31}_{-0.66}$ \\
                \des & $ 0.907^{+0.093}_{-0.081}$ & $ 0.737^{+0.072}_{-0.080}$ & $ 0.678^{+0.081}_{-0.14}$ & $ 1.20^{+0.49}_{-1.1}$ \\
                \southd  & $ 0.909\pm 0.085$ & $ 0.745\pm 0.074$ & $ 0.677^{+0.085}_{-0.096}$ & $ 0.606^{+0.070}_{-0.080}$ \\
                \hline
                \northd No $\shear$ & $ 0.835\pm 0.086$ & $ 0.767\pm 0.056$ & $ 0.658^{+0.058}_{-0.072}$ & $ 0.569^{+0.059}_{-0.069}$ \\
                \northd No $\delta_g$ & $ 0.76^{+0.13}_{-0.17}$ & $ 0.74^{+0.12}_{-0.14}$ & $ 1.04\pm 0.22$ & $ 0.65^{+0.22}_{-0.65}$ \\
                \dls + \kids & $ 0.803\pm 0.046$ & $ 0.731^{+0.043}_{-0.049}$ & $ 0.739\pm 0.082$ & $ 0.68^{+0.23}_{-0.68}$ \\
                \northd & $ 0.807\pm 0.047$ & $ 0.731^{+0.043}_{-0.048}$ & $ 0.719^{+0.058}_{-0.068}$ & $ 0.568\pm 0.065$ \\
                \hline 
                \alld No $\shear$ & $0.876\pm 0.084$ & $ 0.797\pm 0.056$ & $ 0.702^{+0.061}_{-0.068}$ & $ 0.590^{+0.060}_{-0.069}$ \\
                \alld No $\delta_g$ & $ 0.80^{+0.10}_{-0.12}$ & $ 0.720^{+0.083}_{-0.10}$ & $ 0.80\pm 0.15$ & $ 0.61^{+0.16}_{-0.61}$ \\
                \alld & $0.823\pm 0.045$ & $ 0.723\pm 0.042$ & $ 0.718^{+0.053}_{-0.062}$ & $ 0.569^{+0.061}_{-0.069}$\\
                \hline
                \planck 18 \lcdm &$0.8801 \pm 0.0013$ & $0.7571 \pm 0.0021$ & $0.6549 \pm 0.0024$ & $0.4958 \pm 0.0023$\\
                \hline
            \end{tabular}
            \caption{Constraints on the growth reconstruction nodes $\tilde{D}_z$ for different combinations of data sets. Note that a fifth node $\tilde{D}_5 = 0.212$ was used as anchor to match \planck at $z\geq5$ when doing the interpolation. The \planck 18 \lcdm row was obtained reanalyzing the official chains~\cite{1807.06209}.}
            \label{tab:dpk}
        \end{table}

      To further explore the complementarity between different probes, Figure~\ref{fig:DES_dpk} shows the two-dimensional confidence intervals for the growth reconstruction parameters $\tilde{D}_z$ in the case of the \southd experiments for different data set combinations. Besides the complementarity between shear and clustering data sets in terms of the final statistical uncertainties on $\tilde{D}_z$ we just described, we can also interpret the level of correlation between different parameters. While the constraints from the combination of clustering and CMB lensing are approximately uncorrelated between different growth parameters, the constraints from cosmic shear show a clear positive correlation between adjacent $\tilde{D}_z$s. This is also easy to understand in terms of the radial kernel associated to the different tracers. While galaxy clustering traces the amplitude of inhomogeneities locally (i.e. in the redshift range where the galaxies are selected), weak lensing traces the cumulative distribution, and thus the measurements of the shear power spectrum in different redshift bins are visibly correlated.
      
      The final quantitative constraints on $\tilde{D}_z$ for the different data set combinations are listed in Table \ref{tab:dpk}, together with the expected values of the growth factor estimated from the \planck 18 \lcdm  chains \cite{1807.06209}. As is evident from the previous figures, the evidence of tension with respect to CMB data is concentrated on the first two redshift nodes. We find only a mild improvement in the goodness of fit between \lcdm and this more flexible model. For instance, when using the \alld data we find $\Delta\chi^2=-7$, which does not represent strong evidence against \lcdm given the 3 additional parameters of this more flexible model. We therefore conclude that the linear growth history preferred by current large-scale structure data is in good agreement with the expectation from \lcdm \footnote{However, we cannot rule out a Einstein-de Sitter ($\Om=1$) universe with more than $3\sigma$ from our measurements of $\tilde{D}_{z}$ alone.}, and that the main source of tension with CMB data is in the overall amplitude of fluctuations. In the future, it will be interesting to incorporate RSD measurements in this formalism to simultaneously constrain the growth factor and its derivative with respect to redshift.

  \section{Discussion}\label{sec:disc}
    In this paper we set out to answer the question: \emph{``What do current large-scale structure data say about the growth of density fluctuations in the late Universe?''}. To do so, we have carried out a combined analysis of 6 different projected probes of the large-scale structure, separating the linear growth from the background expansion and parametrizing the former in terms of its value at a set of four redshift nodes covering the range $0.2\lesssim z\lesssim 2$. The main motivation for this exercise is to shed light on the tension in the value of the current amplitude of linear fluctuations as measured by low-redshift weak lensing data~\cite{1708.01538, 2007.15632, 1809.09148} and as extrapolated from CMB data in the context of \lcdm. Answering this question has allowed us to a) evaluate the consistency between different low-redshift probes in their predicted growth history and b) determine at what point in time we have evidence of a slower growth than that predicted by \planck.

    Our work is based on an independent harmonic-space-based analysis, making use of state-of-the-art methods to obtain unbiased and precise measurements of power spectra  for different probes and the associated covariance matrix. This has allowed us to carry out a consistent analysis of galaxy clustering from \des, \dls and \eboss, cosmic shear from \kids and \des, and CMB lensing from \planck. The methods used have been extensively validated and, as we have shown, we find no significant evidence of unaccounted systematic uncertainties in our data vector and covariances within the range of scales we use.

    Our main result, shown in Figure \ref{fig:all_s8z}, is the reconstructed growth history at late times. Our analysis has shown that the tension in the measured value of $S_8$ is due to the data preferring a lower amplitude of fluctuations in the range $0.25\lesssim z\lesssim 0.7$ than the \planck prediction with a significance higher than $2\sigma$. The data used here are not able to place strong constraints at lower or higher redshifts and thus show no deviation with respect to the \planck preferred value in this regime. As we have shown, this tension is driven by the cosmic shear data sets, whereas the combination of galaxy clustering and CMB lensing used here shows no significant deviation with respect to \planck. Our uncertainties are not small enough to reveal any tension between these probes, however, although the results found by other groups with different data \citep{2105.03421} suggest that the tension is also revealed by CMB lensing tomography.

    In order to validate our analysis pipeline, we have verified that the constraints on \lcdm from our data agree well with those found by the different collaborations. Doing so has also allowed us to explore the constraints our full complement of data sets is able to place on this model when combined, measuring the $S_8$ parameter to be $S_8=0.7781\pm 0.0094$. This is in good agreement with current constraints from other groups, and in tension with \planck at the $\sim3.4\sigma$ level. This also shows that, in combination, current large-scale structure data are able to constrain these parameters with an uncertainty that is significantly ($\sim25\%$) better than that achieved by primary CMB data.

    The results presented here are subject to some caveats. First, the specific properties of the method used to parametrize the linear growth as a free function of redshift (a quadratic spline), has an impact on the results in terms of e.g. the correlation length of this function, or its behaviour outside of the range of redshifts constrained by the data. A more ideal choice would have been, for example, to make use of Gaussian process regression, in order to let the data constrain the smoothness of this function, and to robustly quantify its uncertainty in poorly-constrained regimes. This approach would significantly increase the complexity of the model and most likely require the use of differentiable sampling methods, and therefore we leave this for future work. 
    
    Since the focus of this paper is not to derive constraints on \lcdm, our analysis of this model has overlooked a number of issues that should be explored in more detail. Although the \northd and \southd data sets are spatially disjoint, the systematic uncertainties in their redshift distribution are, to some extent, correlated due to the use of similar spectroscopic calibration samples. Furthermore, as shown in \cite{1906.09262}, the potential biases in these calibrating samples could give rise to systematic effects in the derived redshift distributions that are not sufficiently well captured by the mean shifts. This is in addition to other unknown observational systematics correlated across overlapping probes. We have ignored the impact of baryons on our final constraints. Although this effect was found to be subdominant within the scales used here by individual collaborations, the increased sensitivity of their combination could make it more relevant. The same can be said about the level of complexity used to parametrize the impact of intrinsic alignments. Finally, our analysis has made use of the simplest linear bias model to describe galaxy clustering, which significantly limits the range of scales over which this tracer can be used. As shown in \cite{2103.09820} the use of recent advances in the modelling of galaxy bias (e.g. \cite{1910.07097,2101.11014}) would allow us to significantly increase this range, which would have a direct impact on the final cosmological constraints. This could be particularly relevant in the case of growth reconstruction, since the galaxy clustering samples act as redshift anchors enabling a more precise recovery of the redshift dependence.

    At this point it is worth asking ourselves what possible explanations could resolve this tension in the light of the growth reconstruction results presented here. The three possible scenarios are the presence of unknown systematic uncertainties in the CMB analysis, the presence of systematic uncertainties in the analysis of cosmic shear (which, as we have shown, is the driving source of this tension), or the presence of new physics giving rise to a slower growth of perturbations at late times. The results found by the Atacama Cosmology Telescope (ACTPol) collaboration independently of \planck data \cite{2007.07288}, obtain $S_8=0.840\pm0.030$ in combination with WMAP data, in perfect agreement with \planck and in $2\sigma$ tension with our results. Therefore there is currently no significant evidence of systematic uncertainties which have been unaccounted for in the \planck measurement. As we have shown, both \des and \kids independently recover the same lower amplitude of fluctuations at $z\sim0.4$, and therefore the possibility of experiment-specific systematic uncertainties being the cause of the tension is not strongly favoured. The other possibility would be a mis-modelling of unknown astrophysical systematic uncertainties specific to cosmic shear and therefore common to different experiments. However, similar results have been found by others without cosmic shear data \cite{2105.03421}.

    We must therefore turn to the third scenario: new physics. A number of extensions to \lcdm have been explored in \cite{1810.02499,2010.16416} in the context of \des and \kids respectively. While some simple extensions, such as allowing for departures of the dark energy equation of state from a perfect cosmological constant, are able to reconcile the measured values of $S_8$, significant tension still remains in the regions of the multi-dimensional parameter space preferred by the CMB and large-scale structure data sets~\cite{2010.16416}. Furthermore, the data are not yet able to detect departures from a flat Universe, as well as any signatures of modified gravity \cite{2010.15278}. A less exotic proposal (see e.g. \cite{2010.00466,2103.09820,2105.03421}) is the possibility of a value of $\Om$ slightly lower than that found by CMB experiments, which would also agree with other probes of structure growth.

    A final scenario, although arguably a less interesting one, is that the current tension is simply a statistical fluke which sits at the $\sim3\sigma$ level for this particular combination of parameters due to a slight mis-estimation of subdominant systematic uncertainties. Fortunately, current constraints are far from sample-variance dominated. Most immediately, the imminent release of the \des third-year analysis is expected to shed light on this issue, either reinforcing or alleviating the existing tension. Further down the line, data from next-generation photometric surveys such as the Vera C. Rubin Observatory \cite{0912.0201} or Euclid \cite{1001.0061}, spectroscopic surveys like DESI \cite{1611.00036}, and CMB experiments such as the Simons Observatory \cite{1808.07445} and CMB Stage-4 \cite{1610.02743} will allow us to resolve and exploit the source of this tension. These data will also allow us to map the growth history in a model-independent way with a much higher precision than current data allow over a larger range of redshifts. The question of whether these forthcoming data will be able to strengthen or alleviate the $S_8$ tension remains open. 
    
    Significant progress can still be made on the constraints presented here with currently-available data. At low redshifts, where the data used here are not able to strongly constrain growth, our measurements could be improved by including information from all-sky photometric surveys \citep{1311.5246,1607.01182}, and, potentially, from maps of the thermal Sunyaev-Zel'dovich effect as a probe of structure \cite{1502.01596,1712.00788,1909.09102,2102.07701} assuming astrophysical uncertainties can be kept under control \citep{1312.5341,2010.07797}. At redshifts $z\gtrsim1$ further information could be gained by combining these data with infrared data such as unWISE \cite{1909.07412,2105.03421} or radio continuum data \cite{1908.10309,2009.01817}. Finally, other high-precision probes of growth, such as redshift-space distortions and peculiar velocity surveys \cite{1609.08247,2105.05185}, should allow us to constrain the growth history at least at the same level as tomographic data, and therefore it is worth developing methods to combine these measurements in a consistent way.

  \section*{Acknowledgements}
    We would like to thank Erminia Calabrese, Boryana Hadzhyiska, Ian Harrison, Shahab Joudaki, An\v{z}e Slosar for useful discussions. We would also like to thank the anonymous referee for their careful reading and useful comments that helped to improve the quality of this work. AN is supported by NSF grant AST-181497. CGG, EB, EMM and PGF are supported by European Research Council Grant No:  693024 and the Beecroft Trust. CGG was also supported by the grant PGC2018-095157-B-I00 from Ministry of Science, Innovation and Universities of Spain and by the Spanish grant, partially funded by the ESF, BES-2016-077038. PRL is supported by the grant PGC2018-095157-B-I00 from Ministry of Science, Innovation and Universities of Spain. DA is supported by the Science and Technology Facilities Council through an Ernest Rutherford Fellowship, grant reference ST/P004474. We also made extensive use of computational resources at the University of Oxford Department of Physics, funded by the John Fell Oxford University Press Research Fund.

    We made extensive use of the {\tt numpy} \citep{oliphant2006guide, van2011numpy}, {\tt scipy} \citep{2020SciPy-NMeth}, {\tt astropy} \citep{1307.6212, 1801.02634}, {\tt healpy} \citep{Zonca2019}, and {\tt matplotlib} \citep{Hunter:2007} python packages.
   
    This paper makes use of software developed for the Large Synoptic Survey Telescope. We thank the LSST Project for making their code available as free software at \url{http://dm.lsst.org}.
   
    This project used public archival data from the Dark Energy Survey (DES). Funding for the DES Projects has been provided by the U.S. Department of Energy, the U.S. National Science Foundation, the Ministry of Science and Education of Spain, the Science and Technology Facilities Council of the United Kingdom, the Higher Education Funding Council for England, the National Center for Supercomputing Applications at the University of Illinois at Urbana-Champaign, the Kavli Institute of Cosmological Physics at the University of Chicago, the Center for Cosmology and Astro-Particle Physics at the Ohio State University, the Mitchell Institute for Fundamental Physics and Astronomy at Texas A\&M University, Financiadora de Estudos e Projetos, Funda{\c c}{\~a}o Carlos Chagas Filho de Amparo {\`a} Pesquisa do Estado do Rio de Janeiro, Conselho Nacional de Desenvolvimento Cient{\'i}fico e Tecnol{\'o}gico and the Minist{\'e}rio da Ci{\^e}ncia, Tecnologia e Inova{\c c}{\~a}o, the Deutsche Forschungsgemeinschaft, and the Collaborating Institutions in the Dark Energy Survey.
        
    The Collaborating Institutions are Argonne National Laboratory, the University of California at Santa Cruz, the University of Cambridge, Centro de Investigaciones Energ{\'e}ticas, Medioambientales y Tecnol{\'o}gicas-Madrid, the University of Chicago, University College London, the DES-Brazil Consortium, the University of Edinburgh, the Eidgen{\"o}ssische Technische Hochschule (ETH) Z{\"u}rich,  Fermi National Accelerator Laboratory, the University of Illinois at Urbana-Champaign, the Institut de Ci{\`e}ncies de l'Espai (IEEC/CSIC), the Institut de F{\'i}sica d'Altes Energies, Lawrence Berkeley National Laboratory, the Ludwig-Maximilians Universit{\"a}t M{\"u}nchen and the associated Excellence Cluster Universe, the University of Michigan, the National Optical Astronomy Observatory, the University of Nottingham, The Ohio State University, the OzDES Membership Consortium, the University of Pennsylvania, the University of Portsmouth, SLAC National Accelerator Laboratory, Stanford University, the University of Sussex, and Texas A\&M University.
    
    Based in part on observations at Cerro Tololo Inter-American Observatory, National Optical Astronomy Observatory, which is operated by the Association of Universities for Research in Astronomy (AURA) under a cooperative agreement with the National Science Foundation.
    
    Based on observations made with ESO Telescopes at the La Silla Paranal Observatory under programme IDs 177.A-3016, 177.A-3017, 177.A-3018 and 179.A-2004, and on data products produced by the KiDS consortium. The KiDS production team acknowledges support from: Deutsche Forschungsgemeinschaft, ERC, NOVA and NWO-M grants; Target; the University of Padova, and the University Federico II (Naples).
    
    The Legacy Surveys consist of three individual and complementary projects: the Dark Energy Camera Legacy Survey (DECaLS; Proposal ID \#2014B-0404; PIs: David Schlegel and Arjun Dey), the Beijing-Arizona Sky Survey (BASS; NOAO Prop. ID \#2015A-0801; PIs: Zhou Xu and Xiaohui Fan), and the Mayall z-band Legacy Survey (MzLS; Prop. ID \#2016A-0453; PI: Arjun Dey). DECaLS, BASS and MzLS together include data obtained, respectively, at the Blanco telescope, Cerro Tololo Inter-American Observatory, NSF's NOIRLab; the Bok telescope, Steward Observatory, University of Arizona; and the Mayall telescope, Kitt Peak National Observatory, NOIRLab. The Legacy Surveys project is honored to be permitted to conduct astronomical research on Iolkam Du\'ag (Kitt Peak), a mountain with particular significance to the Tohono O’odham Nation.

    NOIRLab is operated by the Association of Universities for Research in Astronomy (AURA) under a cooperative agreement with the National Science Foundation.
    
    This project used data obtained with the Dark Energy Camera (DECam), which was constructed by the Dark Energy Survey (DES) collaboration. Funding for the DES Projects has been provided by the U.S. Department of Energy, the U.S. National Science Foundation, the Ministry of Science and Education of Spain, the Science and Technology Facilities Council of the United Kingdom, the Higher Education Funding Council for England, the National Center for Supercomputing Applications at the University of Illinois at Urbana-Champaign, the Kavli Institute of Cosmological Physics at the University of Chicago, Center for Cosmology and Astro-Particle Physics at the Ohio State University, the Mitchell Institute for Fundamental Physics and Astronomy at Texas A\&M University, Financiadora de Estudos e Projetos, Fundacao Carlos Chagas Filho de Amparo, Financiadora de Estudos e Projetos, Fundacao Carlos Chagas Filho de Amparo a Pesquisa do Estado do Rio de Janeiro, Conselho Nacional de Desenvolvimento Cientifico e Tecnologico and the Ministerio da Ciencia, Tecnologia e Inovacao, the Deutsche Forschungsgemeinschaft and the Collaborating Institutions in the Dark Energy Survey. The Collaborating Institutions are Argonne National Laboratory, the University of California at Santa Cruz, the University of Cambridge, Centro de Investigaciones Energeticas, Medioambientales y Tecnologicas-Madrid, the University of Chicago, University College London, the DES-Brazil Consortium, the University of Edinburgh, the Eidgenossische Technische Hochschule (ETH) Zurich, Fermi National Accelerator Laboratory, the University of Illinois at Urbana-Champaign, the Institut de Ciencies de l'Espai (IEEC/CSIC), the Institut de Fisica d'Altes Energies, Lawrence Berkeley National Laboratory, the Ludwig Maximilians Universitat Munchen and the associated Excellence Cluster Universe, the University of Michigan, NSF's NOIRLab, the University of Nottingham, the Ohio State University, the University of Pennsylvania, the University of Portsmouth, SLAC National Accelerator Laboratory, Stanford University, the University of Sussex, and Texas A\&M University.
    
    BASS is a key project of the Telescope Access Program (TAP), which has been funded by the National Astronomical Observatories of China, the Chinese Academy of Sciences (the Strategic Priority Research Program “The Emergence of Cosmological Structures” Grant \# XDB09000000), and the Special Fund for Astronomy from the Ministry of Finance. The BASS is also supported by the External Cooperation Program of Chinese Academy of Sciences (Grant \# 114A11KYSB20160057), and Chinese National Natural Science Foundation (Grant \# 11433005).
    
    The Legacy Survey team makes use of data products from the Near-Earth Object Wide-field Infrared Survey Explorer (NEOWISE), which is a project of the Jet Propulsion Laboratory/California Institute of Technology. NEOWISE is funded by the National Aeronautics and Space Administration.
    
    The Legacy Surveys imaging of the DESI footprint is supported by the Director, Office of Science, Office of High Energy Physics of the U.S. Department of Energy under Contract No. DE-AC02-05CH1123, by the National Energy Research Scientific Computing Center, a DOE Office of Science User Facility under the same contract; and by the U.S. National Science Foundation, Division of Astronomical Sciences under Contract No. AST-0950945 to NOAO.
    
    Funding for the Sloan Digital Sky Survey IV has been provided by the Alfred P. Sloan Foundation, the U.S. Department of Energy Office of Science, and the Participating Institutions. 
    
    SDSS-IV acknowledges support and resources from the Center for High Performance Computing  at the University of Utah. The SDSS website is www.sdss.org.
    
    SDSS-IV is managed by the Astrophysical Research Consortium for the Participating Institutions of the SDSS Collaboration including the Brazilian Participation Group, the Carnegie Institution for Science, Carnegie Mellon University, Center for Astrophysics | Harvard \& Smithsonian, the Chilean Participation Group, the French Participation Group, Instituto de Astrof\'isica de Canarias, The Johns Hopkins University, Kavli Institute for the Physics and Mathematics of the Universe (IPMU) / University of Tokyo, the Korean Participation Group, Lawrence Berkeley National Laboratory, Leibniz Institut f\"ur Astrophysik Potsdam (AIP),  Max-Planck-Institut f\"ur Astronomie (MPIA Heidelberg), Max-Planck-Institut f\"ur Astrophysik (MPA Garching), Max-Planck-Institut f\"ur Extraterrestrische Physik (MPE), National Astronomical Observatories of China, New Mexico State University, New York University, University of Notre Dame, Observat\'ario Nacional / MCTI, The Ohio State University, Pennsylvania State University, Shanghai Astronomical Observatory, United Kingdom Participation Group, Universidad Nacional Aut\'onoma de M\'exico, University of Arizona, University of Colorado Boulder, University of Oxford, University of Portsmouth, University of Utah, University of Virginia, University of Washington, University of Wisconsin, Vanderbilt University, and Yale University. 
    
  \bibliography{main,non_ads}
\appendix
    
    \section{Effect of non-Gaussian covariances}\label{app:cov}
  
      \begin{figure}
        \centering
        \includegraphics[width=0.8\textwidth]{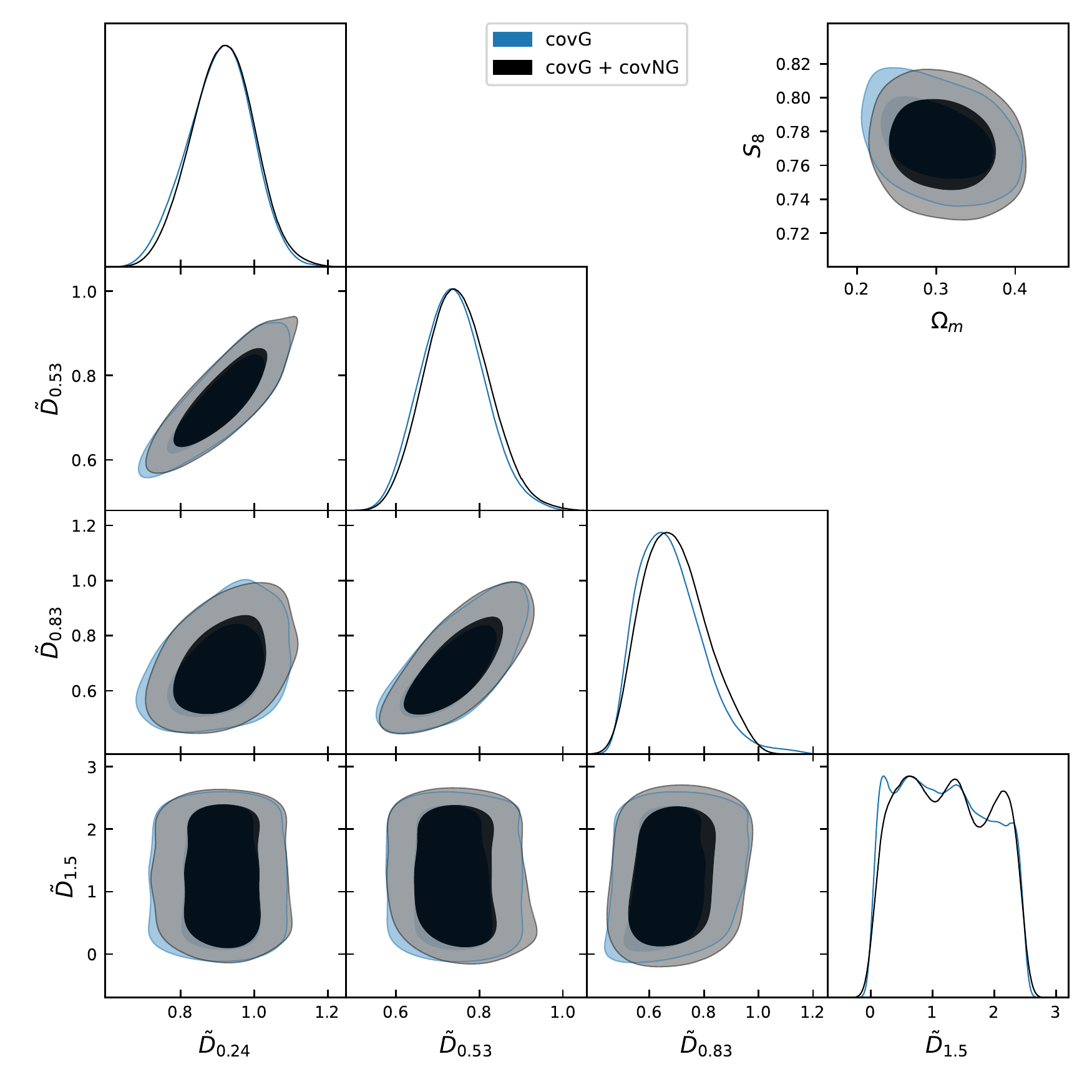}
        \caption{Posterior distribution for the growth reconstruction nodes for the cases using only the Gaussian part of the covariance matrix and the additional non-Gaussian terms. We use only \des data as a worst case scenario where the last node remains unconstrained. On the top right corner, we show the constraints for the $(\Om, S_8)$ parameters for \lcdm using \des + \cmbk. As can be seen, the non-Gaussian part does not affect the posterior distributions. The small differences are likely to be originated by different chain lengths.}\label{fig:covGvscovGNG}
      \end{figure}
          In order to verify that our results are insensitive to the non-Gaussian contributions to the covariance matrix over the range of scales used here, we have repeated the analysis for both \lcdm and the growth reconstruction in two different ways that cover different situations. In the first case, we use the \des $5\times2$pt data including the connected trispectrum and super-sample terms computed as described in \cite{1912.08209}. In comparison, for the reconstruction of growth we only use the \des $3\times2$pt data as a worse case scenario, where there is no cosmological information to constrain the last node $\tilde{D}_{1.5}$.. The results are shown in Figure \ref{fig:covGvscovGNG} for the growth rate parameters $\tilde{D}_z$ and the \lcdm parameters $(\Om,S_8)$. The non-Gaussian contributions have a negligible effect on the final best-fit parameters and their uncertainties within the range of scales used here. The impact on the goodness of fit is also small, with the best-fit $\chi^2$ changing by less than 1\% in both cases. Given these results and for simplicity, we use only the dominant Gaussian part of the covariance matrix, calculated as described in \ref{ssec:meth.cov}.

\end{document}